\documentstyle[aps,pre,twocolumn,floats,epsfig]{revtex}
\begin{document}

\newcommand{\tbox}[1]{\mbox{\tiny #1}} 
\newcommand{\half}{\mbox{\small $\frac{1}{2}$}} 
\newcommand{\mbf}[1]{{\mathbf #1}}


\title{Parametric Evolution for a Deformed Cavity}

\author{Doron Cohen, Alex Barnett and Eric J. Heller}

\date{July 2000}

\address{Department of Physics, Harvard University} 

\maketitle


\begin{abstract} 
We consider a classically chaotic system that is described 
by a Hamiltonian ${\cal H}(Q,P;x)$, where $(Q,P)$ describes 
a particle moving inside a cavity, and $x$ controls  
a deformation of the boundary. 
The quantum-eigenstates of the system are $|n(x)\rangle$.  
We describe how the parametric kernel 
$P(n|m)= |\langle n(x)|m(x_0)\rangle|^2$, 
also known as the local density of states,  
evolves as a function of $\delta x=x{-}x_0$.  
We illuminate the non-unitary nature of this parametric 
evolution, the emergence of non-perturbative features, 
the final non-universal saturation, and the limitations 
of random-wave considerations.  
The parametric evolution is demonstrated numerically 
for two distinct representative deformation processes.    
\end{abstract}

\section{Introduction}

\subsection{The local density of states}

Consider a system that is described by an Hamiltonian
${\cal H}(Q,P;x)$ where $(Q,P)$ are canonical variables and
$x$ is a constant parameter. Our interest in this paper 
is in the case where $(Q,P)$ describe the motion of 
a particle inside a cavity, and $x$ controls the deformation 
of the confining boundary. The 1D version of a cavity, 
also known as `potential well', is illustrated in Fig.1. 
However, we are mainly interested in the case of chaotic 
cavities in $d>1$ dimensions. Cavities in $d=2$ dimensions, 
also known as billiard systems, are prototype examples 
in the studies of classical and quantum chaos, and we shall 
use them for the purpose of numerical illustrations.

\begin{figure}[hb]
\centerline{\epsfig{figure=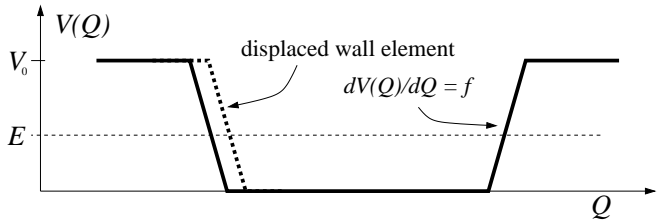,width=\hsize}}
\vspace{.1in}
\caption{
The shape of a cavity in $d$ dimensions 
is defined by its $d-1$ boundary. 
The confining potential is $V(Q)$. 
The figure illustrates $V(Q)$ for 1D well. 
It also can be regarded as a cross section of 
$d>1$ cavity. 
The kinetic energy of the particle is $E= \half mv^2$.  
The walls of the cavity exert a field of force $f$ 
on the bouncing particle. The hard wall 
limit corresponds to $f\rightarrow\infty$ 
and $V_0\rightarrow\infty$.  
For theoretical considerations it is convenient 
to assume that $f$ and $V_0$ are large but finite. 
Mathematically it is also convenient to think 
of the embedding space as having some huge but 
finite volume (not illustrated).} 
\end{figure}

The eigenstates of the quantized Hamiltonian are $|n(x)\rangle$ 
and the corresponding eigen-energies are $E_n(x)$.
The eigen-energies are assumed to be ordered, and the
mean level spacing will be denoted by $\Delta$.
We are interested in the parametric kernel
\begin{eqnarray} \label{e1}
P(n|m) \ = \ |\langle n(x)|m(x_0)\rangle|^2
\ = \ \mbox{trace}(\rho_n\rho_m)
\end{eqnarray}
In the equation above $\rho_m(Q,P)$ and $\rho_n(Q,P)$
are the Wigner functions that correspond to the
eigenstates $|m(x_0)\rangle$ and $|n(x)\rangle$
respectively. The trace stands for $dQdP/(2\pi\hbar)^d$
integration. The difference $x-x_0$ will be denoted by $\delta x$.
We assume a dense spectrum, so that our interest is in 
`classically small' but `quantum mechanically large' 
energy scales. It is important to realize 
that the kernel $P(n|m)$ has a well defined classical limit. 
The classical approximation (see remark \cite{rmrk1})  
is obtained by using microcanonical distributions instead 
of Wigner functions.

Fixing $n$, the vector $P(n|m)$ describes 
the shape of the $n$-th eigenstate in the 
${\cal H}_0 = {\cal H}(Q,P;x_0)$ representation. 
By averaging over several eigenstates one obtains the 
average shape of the eigenstate (ASOE).  
We can also identify $P(n|m)$ as the 
local density of states (LDOS), 
by regarding it as a function of $n$, 
where $m$ is considered to be a fixed 
reference state. In the latter case 
an average over few $m$-states is assumed. 
We shall denote the LDOS by $P(r)$ where 
$r=(n-m)$. The ASOE is just $P(-r)$. 
Note that the ASOE and the LDOS are 
given by the same function. One would 
have to be more careful with these 
definitions if ${\cal H}_0$ were  
integrable while ${\cal H}$ non-integrable.

A few words are in order regarding the definition 
of the LDOS, and its importance in physical applications. 
The LDOS, also known as strength function \cite{casati,flamb,felix}, 
describes an energy distribution. Conventionally 
it is defined as follows:
\begin{eqnarray} \label{e_2}
\mbox{LDOS}(E) \ \ &=& \ \ 
-\frac{1}{\pi} \ \langle m | \Im \mbf{G}(E) | m \rangle 
\nonumber \\
\ \ &=& \ \ \sum_n P(n|m) \ \delta(E-E_n) 
\end{eqnarray}
where $\mbf{G}(E)=1/(E-{\cal H}+i0)$ is the 
retarded Green function. We are interested in 
chaotic systems, so it should be clear that 
our $P(r)$ is related by trivial change of variable 
($E \mapsto r$) to the above defined $\mbox{LDOS}(E)$. 
Our $P(r)$ also incorporates an average over 
the reference state. The LDOS is important in studies 
of either chaotic or complex conservative quantum 
systems that are encountered in nuclear physics 
as well as in atomic and molecular physics. Related 
applications may be found in mesoscopic physics. 
Going from ${\cal H}_0$ to ${\cal H}$ may signify 
a physical change of an external field, 
or switching on of a perturbation, 
or a sudden-change of an effective-interaction 
(as in molecular dynamics \cite{heller}). 
The so called `line shape' of the LDOS is important 
for the understanding of the associated dynamics. 
It is also important to realize that the LDOS 
is the Fourier transform of the so-called  
`survival probability amplitude' \cite{heller}   
(see \cite{wls} for concise presentation of this point).

\subsection{Parametric Evolution}

Textbook \cite{ct} formulations of 
perturbation theory can be applied in order to find 
the LDOS. Partial summations of diagrams to infinite order 
can be used in order to get an improved 
Lorentzian-type approximation.  However most 
textbooks do not illuminate the limitations  
and the subtleties which are involved in  
using the conventional perturbative schemes. 
It is therefore interesting to take a 
somewhat different approach to the 
study of LDOS. The roots of this alternate approach 
can be traced back to the work of Wigner \cite{wigner} 
regarding a simple banded random matrix (BRM) model 
${\cal H}=\mbf{E}+\delta x\mbf{B}$.  
Here $\mbf{E}$ is a diagonal matrix 
whose elements are the ordered energies $\{E_n\}$, 
and $\mbf{B}$ is a banded matrix. The study of this 
model can be motivated by the realization that 
in generic circumstances it is possible to write 
${\cal H}(Q,P;x) \approx {\cal H}_0(Q,P) + \delta x {\cal F}(Q,P)$. 
Using a simple semiclassical argument \cite{mario} 
it turns out that the matrix representation of any 
generic ${\cal F}$, in the eigen-basis that is 
determined by the chaotic Hamiltonian ${\cal H}_0$,  
is a banded matrix.

The important ingredient (from our point of view) 
in the original work by Wigner, is the emphasis 
on the parametric-evolution (PE) of the LDOS. 
The LDOS describes an energy distribution:  
For $\delta x=0$ the kernel $P(r)$ is simply 
a Kroneker delta function. As $\delta x$ becomes 
larger, the width as well as the whole profile of 
this distribution `evolves'. 
Wigner has realized that for his WBR model there 
are three parametric regimes. For very small 
$\delta x$ we have the standard perturbative 
structure where most of the probability is 
concentrated in $r=0$. For larger $\delta x$ 
we have a Lorentzian line shape. But this 
Lorentzian line shape does not persist if we 
further increase $\delta x$. Instead we get 
a semi-circle line shape. Many works about 
the LDOS have followed \cite{casati,flamb,felix},   
but the issue of PE has not been further 
discussed there. The emphasis in those works 
is mainly on the case where ${\cal H}_0$ is 
an integrable or non-interacting system, while 
${\cal H}$ is possibly (but not necessarily) 
chaotic due to some added perturbation term.

The line of study which is pursued in the present 
work has been originated and motivated by studies 
of quantum dissipation \cite{crs,vrn,frc}. 
Understanding PE can be regarded as a preliminary
stage in the analysis of the energy spreading 
process in driven mesoscopic systems. Note that 
the LDOS gives the energy re-distribution due to 
a `sudden' (very fast) change of the Hamiltonian.
Unlike the common approach for studies of LDOS,  
we assume both ${\cal H}$ and  ${\cal H}_0$ to be 
chaotic. Both correspond to the same 
parametrically dependent Hamiltonian 
${\cal H}(Q,P;x)$, and there is nothing 
special in choosing a particular value $x=x_0$ 
as a starting point for the PE analysis.

\subsection{Main results}

The theory of PE, as discussed 
in general in \cite{crs,vrn,frc} and in 
particular in \cite{wls,lds} takes us beyond 
the random-matrix-theory considerations 
of Wigner. There appear five (rather than two) 
different parametric scales (see remark \cite{rmrk2}). 
These are summarized by Table 1.

\begin{table}[h]
\begin{tabular}{lll}
$\delta x_c^{\tbox{cl}}$  &  Is it possible to linearize ${\cal H}(x+\delta x)$? \\
$\delta x_c^{\tbox{qm}}$  &  Is it possible to use standard perturbation theory?\\
$\delta x_{\tbox{prt}}$   &  Do perturbative tail regions survive? \\
$\delta x_{\tbox{NU}}$    &  Do non-universal core features show up? \\
$\delta x_{\tbox{SC}}$    &  Is it possible to use semiclassical approximation? \\
\end{tabular} 
\vspace{.1in}
\caption{
The parametric scales in the general theory of PE 
are listed (left column) along with the questions 
that motivate their introduction. 
The distribution $P(r)$ may contain perturbative tail regions 
(for $\delta x \ll \delta x_{\tbox{prt}}$),  
and non-perturbative core regions
(for $\delta x > \delta x_c^{\tbox{qm}}$).  
Non-universal (system specific) features may manifest themselves in 
the core structure for $\delta x \gg \delta x_{\tbox{NU}}$.
In generic examples $\delta x \gg \delta x_{\tbox{SC}}$
allows classical approximation for $P(r)$. 
We are going to explain that only two independent 
parametric scales survive in the hard wall limit. }
\end{table}

In the present paper we consider cavities with hard walls.
We are going to explain that because of the 
`hard wall limit' there are only {\em two} 
independent parametric scales:  
One is $\delta x_c^{\tbox{qm}}$ and the others 
(see remark \cite{rmrk3})
coincide with $\delta x_{\tbox{NU}}$.
Assuming that $\delta x_c^{\tbox{qm}}$ and 
$\delta x_{\tbox{NU}}$ are well separated, 
it follows that there are {\em three} distinct 
parametric regimes in the PE of our system. 
These are the standard perturbative regime 
($\delta x \ll \delta x_c^{\tbox{qm}}$), 
the core-tail regime 
($\delta x_c^{\tbox{qm}} \ll \delta x \ll \delta x_{\tbox{NU}}$),  
and the non-universal regime 
($\delta x \gg \delta x_{\tbox{NU}}$).

The exploration of the three parametric regimes  
in the PE of a deformed cavity with hard walls is 
the main issue of the present paper. To the best 
of our knowledge such detailed exploration has  
not been practical in the past. We owe our 
ability to carry out this task to a new 
powerful technique for finding clusters 
of billiard eigenstates \cite{vergini,alexthesis}. 
There are also some secondary 
issues that we are going to address:

{\bf (a)} In the strict limit of hard walls the PE becomes 
non-unitarity. We shall use the 1D well example 
in order to shed light on this confusing issue. 
In particular we demonstrate that any truncation 
of the PE equation leads to false unitarity due to 
a finite-size edge effect. 

{\bf (b)} For special deformations, namely those that constitute 
linear combination of translations rotations and dilations, 
the parametric scales $\delta x_c^{\tbox{qm}}$ and 
$\delta x_{\tbox{NU}}$ coincide. Consequently 
there is no longer distinct core-tail regime, and 
the PE features a quite sharp transition from  the 
standard perturbative regime to the non-universal regime. 

{\bf (c)} In the non-universal regime we demonstrate 
that our numerical results are in accordance with 
our theoretical expectation \cite{wls}. 
Namely, the width of the LDOS profile is determined 
by time-domain semiclassical considerations, rather 
then by phase-space or random-wave considerations.  

{\bf (d)}  The last section puts our specific study 
in a larger context. We explain why Wigner's scenario 
of PE is not followed once hard walls are considered.

\section{The cavity system}

We consider a particle moving inside $d$ dimensional cavity 
whose volume is $\mathsf{V}$.   
The kinetic energy of the particle is $E=\half mv^2$, 
where $m$ is its mass, and $v$ is its velocity. 
It is assumed that this motion is classically chaotic.
The ballistic mean free path is $\ell_{\tbox{bl}}$. 
One can use the estimate 
$\ell_{\tbox{bl}} \sim \mathsf{V}/\mathsf{A}$, 
where $\mathsf{A}$ is the total area of the walls.  
The associated time scale is 
$\tau_{\tbox{bl}} = \ell_{\tbox{bl}}/v$.

The penetration distance upon a collision is $\ell=E/f$,
where $f$ is the force that is exerted by the wall.
Upon quantization we have an additional length
scale, which is the De-Broglie wavelength
$\lambda_{\tbox{B}}=2\pi\hbar/(mv)$.  We shall distinguish
between the {\em hard walls} case where we assume
$\ell < \lambda_{\tbox{B}} \ll \ell_{\tbox{bl}}$,
and {\em soft walls} for which  $\lambda_{\tbox{B}} \ll \ell$.
Note that taking $\hbar\rightarrow 0$ implies soft walls.

There is a class of special deformations that are 
shape-preserving. These are generated by translations, 
rotations and dilations of the cavity.   
A general deformation need not preserve the 
billiard shape nor its volume. 
We can specify any deformation by a function $D(\mbf{s})$,
where~$\mbf{s}$ specifies the location of a wall element 
on the boundary (surface) of the cavity, 
and $D(\mbf{s})\delta x$ is the normal displacement of 
this wall element. 
In many practical cases it is possible to use the convention 
$|D(\mbf{s})| \sim 1$. With this convention $\delta x$ has 
units of length, and its value has the meaning of typical 
wall displacement.

The eigen-energies of a particle inside the cavity 
are in general $x$-dependent, and can be written 
as  $E_n = (\hbar k_n)^2/(2m)$.  The mean level spacing is 
\begin{eqnarray} \label{e2}
\Delta = \hbar v \times 
\frac{2\pi}{\Omega_d} 
\ \frac{1}{{\mathsf V}} 
\lambda_{\tbox{B}}^{d{-}1}
\end{eqnarray}
where $\Omega_d=2\pi,4\pi,\cdots$ for $d=2,3,\cdots$.
In our numerical study we shall consider a quarter 
stadium with curved edge of radius $1$ and straight 
edge of length $1$. The `volume' of the quarter stadium
is ${\mathsf V}=1+\pi/4$. The 'area' of its boundary  
${\mathsf A}=4+\pi/2$ is just the perimeter.
We shall look on the parametric evolution of eigenstates around 
$k\sim 400$ where the mean level spacing in $k$ units 
is $\tilde{\Delta}=\Delta/(\hbar v)\approx 0.0088$.

\section{Parametric evolution}

Consider the quantum-mechanical state $\psi=|m(x_0)\rangle$. 
We can write $\psi = \sum_n a_n(x) |n(x)\rangle$. 
The parametric kernel can be written as 
$P(n|m)=|a_n(x)|^2$. It is a standard exercise 
to obtain (from $d\psi/dx=0$ and  differentiating by parts) 
the following equation for the amplitudes: 
\begin{eqnarray} \label{e3} 
\frac{da_n}{dx} \ = \ 
-\frac{i}{\hbar}\sum_{m} \mbf{W}_{nm}(x) \ a_{m}
\end{eqnarray}
In order to get $P(n|m)$ one should solve this 
equation with the initial conditions $a_n(x_0)=\delta_{nm}$. 
The transitions between 
levels are induced by the matrix elements 
\begin{eqnarray} \label{e4} 
{\mathbf W}_{nm} =  
\frac{i\hbar}{E_n{-}E_m} 
\left(\frac{\partial{\cal H}}{\partial x}\right)_{nm} 
\end{eqnarray}
and we use the `gauge' convention 
${\mathbf W}_{nm}{=}0$ for $n{=}m$. 
(Only one parameter is being changed and 
therefore Berry's phase is not an issue).

Eq.(\ref{e3}) is a possible starting point for 
constructing a perturbation theory for the PE of $P(r)$.  
See Ref.\cite{frc} for more details. 
As an input for this equation we need the 
matrix elements of ${\partial{\cal H}}/{\partial x}$. 
These can be calculated using a simple 
boundary integral formula \cite{berry} whose 
simplest derivation \cite{frc} is as follows:  
The position of the particle in the vicinity 
of a wall element is $Q=(z,\mbf{s})$,  
where $\mbf{s}$ is a surface coordinate 
and $z$ is a perpendicular `radial' coordinate. 
We take $f=\infty$ so that 
\begin{eqnarray}  \label{e5} 
\frac{\partial {\cal H}}{\partial x} \ = \ 
- D(\mbf{s}) \ V_0 \ \delta(z)
\end{eqnarray}
The logarithmic derivative of the wavefunction  
on the boundary is $\varphi(\mbf{s})/\psi(\mbf{s})$ where 
$\varphi(\mbf{s})=\mbf{n}{\cdot}\nabla\psi$, and 
$\mbf{n}$ is a unit vector in the $z$ direction.   
For $z>0$ the wavefunction $\psi(Q)$ is a decaying 
exponential. If $V_0$ is large enough, then the 
exponential decay is fast, and we can treat 
the boundary as if it were locally flat. 
It follows that the logarithmic derivative of the wavefunction 
on the boundary should be equal to $-\sqrt{2mV_0}/\hbar$.  
Consequently one obtains the following expression for the 
matrix elements: 
\begin{eqnarray}   \label{e6}
\left(\frac{\partial {\cal H}}{\partial x}\right)_{nm} 
\ = \
-\frac{\hbar^2}{2m}\oint \varphi_n(\mbf{s})\varphi_m(\mbf{s}) 
\ D(\mbf{s}) d\mbf{s}
\end{eqnarray}
In the one-dimensional case the boundary integral 
is replace by the sum $\sum_s \varphi_n(s) \varphi_m(s) D(s) $ 
where $s=1,2$ are the two turning points of the 
potential well.

\section{Hard walls and non-Unitarity}

For the purpose of the following argumentation 
it is convenient to take $f=\infty$, but to 
keep $V_0$ large but finite.  
Mathematically it is also convenient to think 
of the embedding space as having some huge but 
finite volume.  
We would like to illuminate a subtlety which is associated 
with the hard wall limit $V_0\rightarrow\infty$. 
For any finite $V_0$ the parametric kernel satisfies 
\begin{eqnarray} \label{e7} 
\sum_n P(n|m) = p_{\tbox{total}} 
\end{eqnarray}
with $p_{\tbox{total}}=1$. This follows from the 
fact that $|n(x)\rangle$ is a complete orthonormal 
basis for any $x$. However, for hard walls ($V_0=\infty$)  
this statement is not true. This implies that 
for hard walls the PE in non-unitary.  
We are going to explain this point below.

Let us denote the volume of the original cavity 
by ${\mathsf V}_0$ and of the deformed cavity 
by ${\mathsf V}$.  The volume shared by the deformed 
and the undeformed cavities will be denoted by 
${\mathsf V}_0\cap{\mathsf V}$ and we shall use 
the notation $\eta=({\mathsf V}_0\cap{\mathsf V})/{\mathsf V}_0$. 
For the purpose of the following argumentation 
let us consider a reference state $m$ whose energy 
$E_m$ is well below $V_0$. Let us also assume 
that the wall displacement is large compared to De-Broglie 
wavelength. Consequently the expression for 
$P(n|m)$ has the following semiclassical structure: 
\begin{eqnarray} \label{e_7} 
P(n|m) \ \approx \ \eta \times f(E_n{-}E_m) 
+  (1{-}\eta)\times g(E_n{-}E_c)  
\end{eqnarray}
where $E_c = (V_0+E_m)\sim V_0$. The above result
can be deduced by assuming that the wavefunctions 
look ergodic in space, but still that they are characterized 
by a well-defined {\em local} wavelength. 
An equivalent derivation is obtained by using  
the phase-space picture of \cite{wls}.     
Both $f$ and $g$ in the above expression have unit 
normalization, and therefore $p_{\tbox{total}}=1$ 
for any finite  $V_0$. However, for hard walls 
($V_0=\infty$) we have $E_c=\infty$ and therefore 
$p_{\tbox{total}}=\eta$. We may say the the operation 
of taking the hard wall limit does not commute 
with the summation in Eq.(\ref{e7}).   
An analogous statement can be derived regarding 
the summation $\sum_m P(n|m)$, with the respective   
definition $\eta=({\mathsf V}_0\cap{\mathsf V})/{\mathsf V}$.

The correctness of the above observation 
becomes less trivial if we consider (\ref{e3}) 
with expressions (\ref{e4}) and (\ref{e6}) 
substituted for the matrix elements.  
Looking on (\ref{e4}) with (\ref{e6}) it looks 
as if the matrix ${\mathbf W}_{nm}$ is  
hermitian, and therefore should generate 
unitary PE. But this statement is 
mathematically correct only for (any) finite 
truncation $N$ of the PE equation. 
For $N=\infty$ the matrix ${\mathbf W}_{nm}$ 
becomes non-hermitian. It turns out that 
for any finite $N$, there is a pile-up 
of probability in the edges of the spreading 
profile, due to finite-size effect. 
We shall demonstrate this effect in the next 
section using a simple 1D example.  
In other words, if we solve Eq.(\ref{e3}) 
for hard-walled cavity, we get as a result Eq.(\ref{e_7}) 
with $E_c=E_N$.  For $N=\infty$ we get $E_c=\infty$ 
and therefore $p_{\tbox{total}}=\eta$ in accordance 
with the conclusion of the previous paragraph.

Thus if either $V_0<\infty$ or $N<\infty$ then we have 
unitary PE. But for hard walls, meaning 
$V_0=\infty$ with $N=\infty$, we have non-unitary PE.  
The lost probability is associated with the second 
term in Eq.(\ref{e_7}). This term is peaked around 
a high energy $E_c$. For hard walls $E_c=\infty$ and 
consequently some probability is lost.   
The above picture is supported by the simple 
pedagogical example of the next section.

\section{Parametric evolution for a 1D box}

Consider a 1D box with hard walls, 
where the free motion of the 
particle is within $0<Q<{\mathsf a}$. 
The eigenstates of the Hamiltonian are
\begin{eqnarray} \label{e8} 
|n({\mathsf a})\rangle \ \ \longrightarrow \ \ 
(-1)^n \sqrt{\frac{2}{{\mathsf a}}} \ \sin(k_nQ) 
\end{eqnarray}
where $k_n=n\times(\pi/{\mathsf a})$ is the wavenumber, 
and $n=1,2,...$ is the level index. 
The phase factor $(-1)^n$ has been introduced 
for convenience. We consider now the 
parametric evolution as a function of ${\mathsf a}$. 
One easily obtains 
\begin{eqnarray} \label{e9} 
\langle n({\mathsf a}) | m({\mathsf a}_0) \rangle \ = \ 
(-1)^{n} \sqrt{\eta} \  
\frac{\sin(\pi\eta n)}{\pi}
\frac{2m}{\eta^2 n^2 - m^2}
\end{eqnarray}
where $\eta\equiv{\mathsf a}_0/{\mathsf a}$ is assumed to be 
smaller than $1$, corresponding to expansion 
of the box. The probability kernel is 
$P(n|m)=|\langle n | m \rangle|^2$.
One can verify that the parametric 
evolution in the ${\mathsf a}_0\mapsto{\mathsf a}$ direction 
is unitary, meaning that $\sum_n P(n|m) = 1$. 
On the other hand in the ${\mathsf a}\mapsto{\mathsf a}_0$ 
direction the parametric evolution is 
non-unitary, because  $\sum_m P(n|m) = \eta$. 
The profile of $P(n|m)$ for fixed $n$ is illustrated 
by a dashed line in Fig.2.

\begin{figure}
\centerline{\epsfig{figure=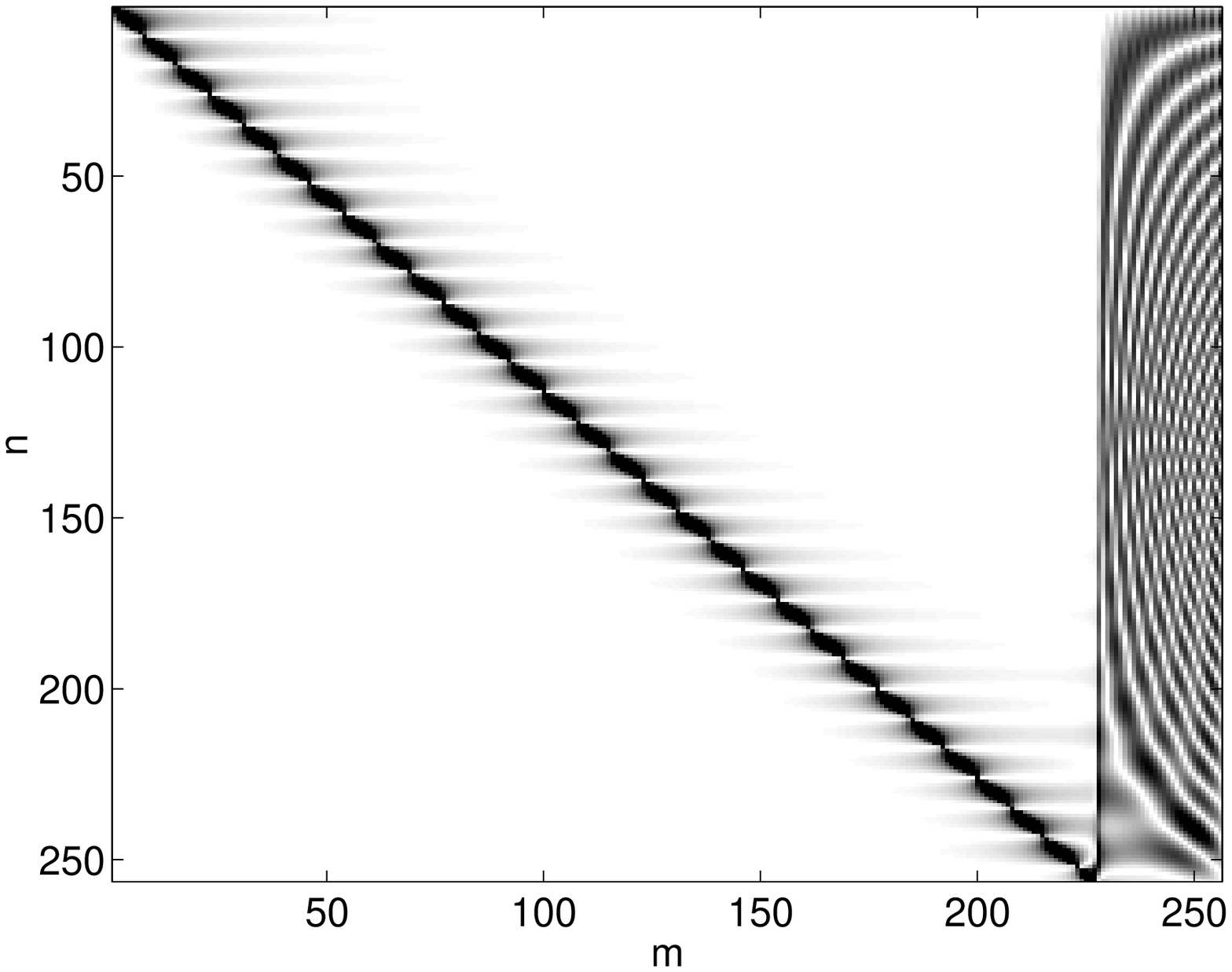,width=\hsize}}
\centerline{\epsfig{figure=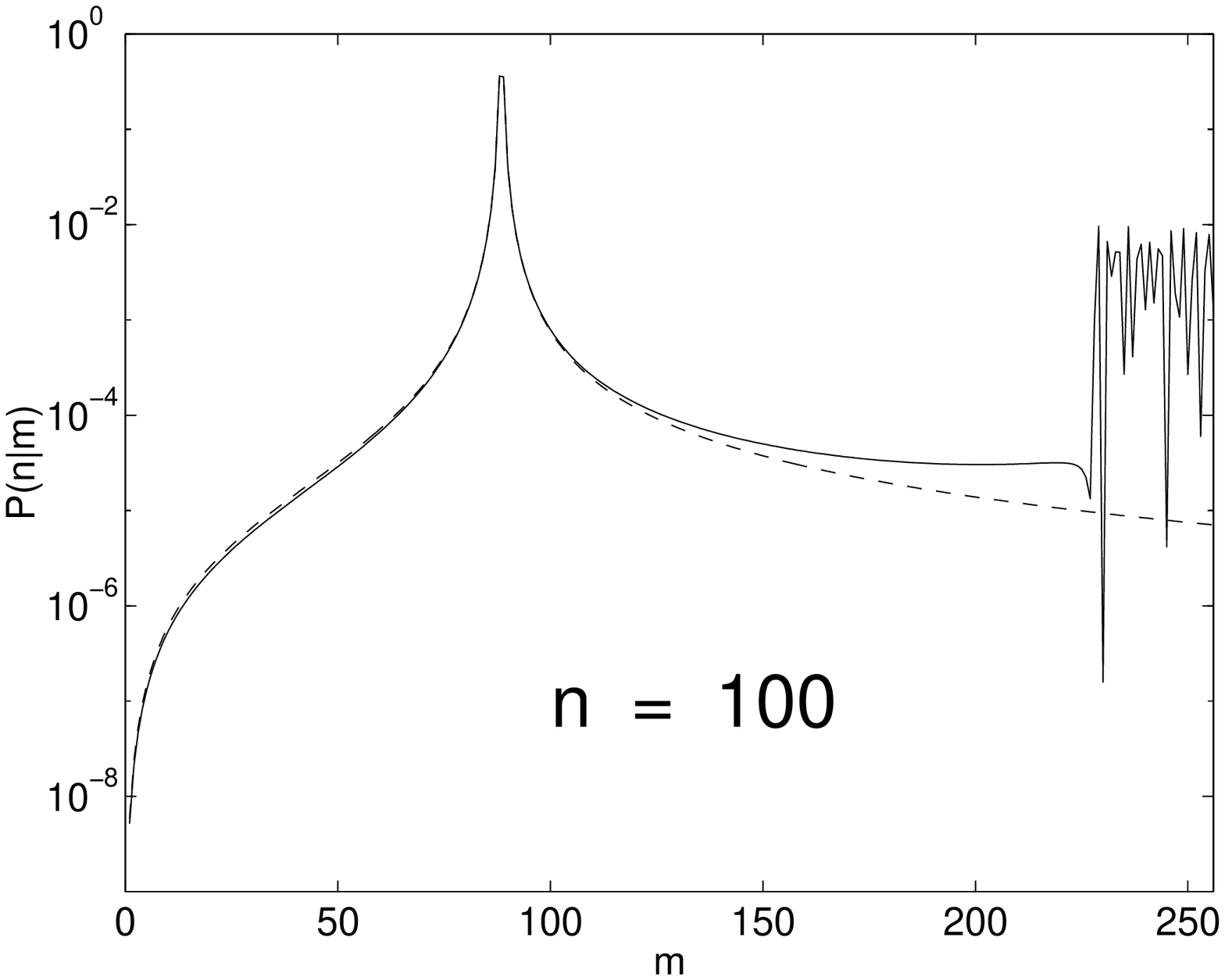,width=\hsize}}
\vspace{.1in}
\caption{
{\em Upper subfigure}: An image of the kernel $P(n|m)$ 
for $13\%$ expansion of the box (ie ${\mathsf a}/{\mathsf a}_0=1.13$). 
The kernel has been calculated numerically using 
Eq.(\ref{e11}) with finite truncation $N=256$. 
{\em Lower subfigure}: The profile of a representative 
row of $P(n|m)$. The dashed line is the $N=\infty$ 
analytical result using Eq.(\ref{e9}).
} 
\end{figure}

We can restore unitarity by making $V_0$ large 
but finite. In such case, a variation of the above 
calculation leads to the following picture: 
Consider the overlap of a reference level $n({\mathsf a})$ 
with the levels $m({\mathsf a}_0)$. As in the 
case $V_0=\infty$ there is a probability $\eta$ 
which is located in the levels whose energies are 
$E_m \approx E_n$. But now the ``lost'' probability 
$(1-\eta)$ is located in the levels 
whose energies are $E_m \approx E_n + V_0$. 
Thus we have $\sum_m P(n|m) = 1$ rather than 
$\sum_m P(n|m) = \eta$.

We consider again the case $V_0=\infty$.  
The normal derivative on the 
boundary is $\varphi_n({\mathsf a})=\sqrt{(2/{\mathsf a})}\ k_n$. 
Hence we can easily get the following result 
\begin{eqnarray}  \label{e10} 
\frac{1}{\hbar}\mbf{W}_{nm} \ = \  
\frac{-i}{k_n^2-k_m^2}\varphi_n({\mathsf a})\varphi_m({\mathsf a}) \ = \ 
-i \frac{1}{{\mathsf a}}\frac{2nm}{n^2-m^2}
\end{eqnarray}
It is more convenient to use 
$\alpha=\ln({\mathsf a})$ for parameterization. 
Hence the equation that describes the 
parametric evolution is 
\begin{eqnarray} \label{e11} 
\frac{da_n}{d\alpha}  \ = \  
-\sum_m \frac{2nm}{n^2-m^2} \ a_m
\end{eqnarray}
For any finite truncation $N<\infty$ this equation 
manifestly generates unitary parametric evolution. 
It is only for $N=\infty$ that it becomes 
equivalent to the non-unitary evolution of 
the 1D box. Again, one can wonder where the 
`lost' probability is located if $N<\infty$.
The answer is illustrated in Fig.2. We see that 
the `lost' probability piles up at the edge 
of the (truncated) tail.

\section{Matrix elements for chaotic cavity}

It is possible to use semiclassical considerations \cite{mario} 
in order to determine the band profile of the matrix Eq.(\ref{e6}).   
The application to the cavity example has been introduced 
in \cite{frc}, and numerically demonstrated in \cite{dil}. 
The accuracy of this semiclassical estimate is remarkable.     
Here we summarize the recipe. First one should generate 
a very long (ergodic) trajectory, 
and define for it the fluctuating quantity 
\begin{eqnarray} \label{e12} 
{\cal F}(t) = -\frac{\partial {\cal H}}{\partial x} = 
\sum_{\tbox{col}} 2mv \ \cos(\theta_{\tbox{col}}) 
\ D_{\tbox{col}} \ \delta(t-t_{\tbox{col}}) 
\end{eqnarray}
where $t_{\tbox{col}}$ is the time of a collision, 
$D_{\tbox{col}}$ stands for $D(\mbf{s})$ 
at the point of the collision, and $v\cos(\theta_{\tbox{col}})$   
is the normal component of the particle's velocity. 
Each delta spike (for soft walls it is actually a narrow 
rectangular spike) corresponds to one collision. 
Now one can calculate the correlation function 
$C(\tau)$ of the fluctuating quantity ${\cal F}(t)$, 
and its Fourier transform $\tilde{C}(\omega)$. 
The semiclassical estimate for the band profile is 
\begin{eqnarray} \label{e13} 
\left\langle\left|\left(
\frac{\partial {\cal H}}{\partial x}\right)_{nm}
\right|^2\right\rangle 
\ \ \approx \ \ 
\frac{\Delta}{2\pi\hbar} \ 
\tilde{C}\left(\frac{E_n{-}E_m}{\hbar}\right)
\end{eqnarray}

Ref.\cite{wlf} contains a systematic study of the 
function $\tilde{C}(\omega)$. For large $\omega$, 
meaning $\omega \gg 1/\tau_{\tbox{bl}}$, one can use  
\begin{eqnarray} \label{e14} 
\tilde{C}(\omega) \ \approx \ 
2 m^2 v^3 \langle|\cos\theta|^3\rangle 
\frac{1}{{\mathsf V}} \oint [D(\mbf{s})]^2 d\mbf{s}
\end{eqnarray}
where the geometric factor is 
$\langle|\cos(\theta)|^3\rangle = 1, 4/3\pi, \cdots$
for $d=1,2, \cdots$. 
A lengthy calculation \cite{frc} reveals that  
Eq.(\ref{e13}) with (\ref{e14}) substituted,  
is an exact global result if we could assume 
that the cavity eigenfunctions look like `random waves', 
and that different wavefunctions are uncorrelated. 
However, it turns out that to take this `random wave' 
result as a global approximation is an over-simplification.  
For $\omega \ll 1/\tau_{\tbox{bl}}$, using the semiclassical 
recipe and assuming strongly chaotic cavity, one obtains  
\begin{eqnarray} \label{e15} 
\tilde{C}(\omega) \ \approx \ 
\tilde{C}(\infty) \times  (\tau_{\tbox{bl}} \omega)^{\gamma}
\end{eqnarray}
with $\gamma=4$ for dilations and translations, 
$\gamma=2$ for rotations, and $\gamma=0$ for 
normal deformations. We use the term 
'special deformations' \cite{wlf} in order 
to distinguish those deformations that has 
the property $\tilde{C}(\omega) \rightarrow 0$ 
in the limit $\omega\rightarrow0$. 
Any combination of dilations, translations and 
rotations is a special deformation. 
Around the bouncing frequency ($\omega \sim 1/\tau_{\tbox{bl}}$)  
the function $\tilde{C}(\omega)$ typically displays 
some non-universal (system and deformation specific) structure.  
This is true for any typical deformation, but for some 
deformations the non-universal features are more pronounced.  
If the cavity has bouncing ball modes, we may get also a modified  
(non-universal) behavior in the small frequency limit. 
For the purpose of general discussion 
it is convenient to assume that the interpolation 
between (\ref{e15}) and (\ref{e14}) is 
smooth, but in actual numerical calculation the 
actual $\tilde{C}(\omega)$ is computed (see below).

\begin{figure}
\centerline{\epsfig{figure=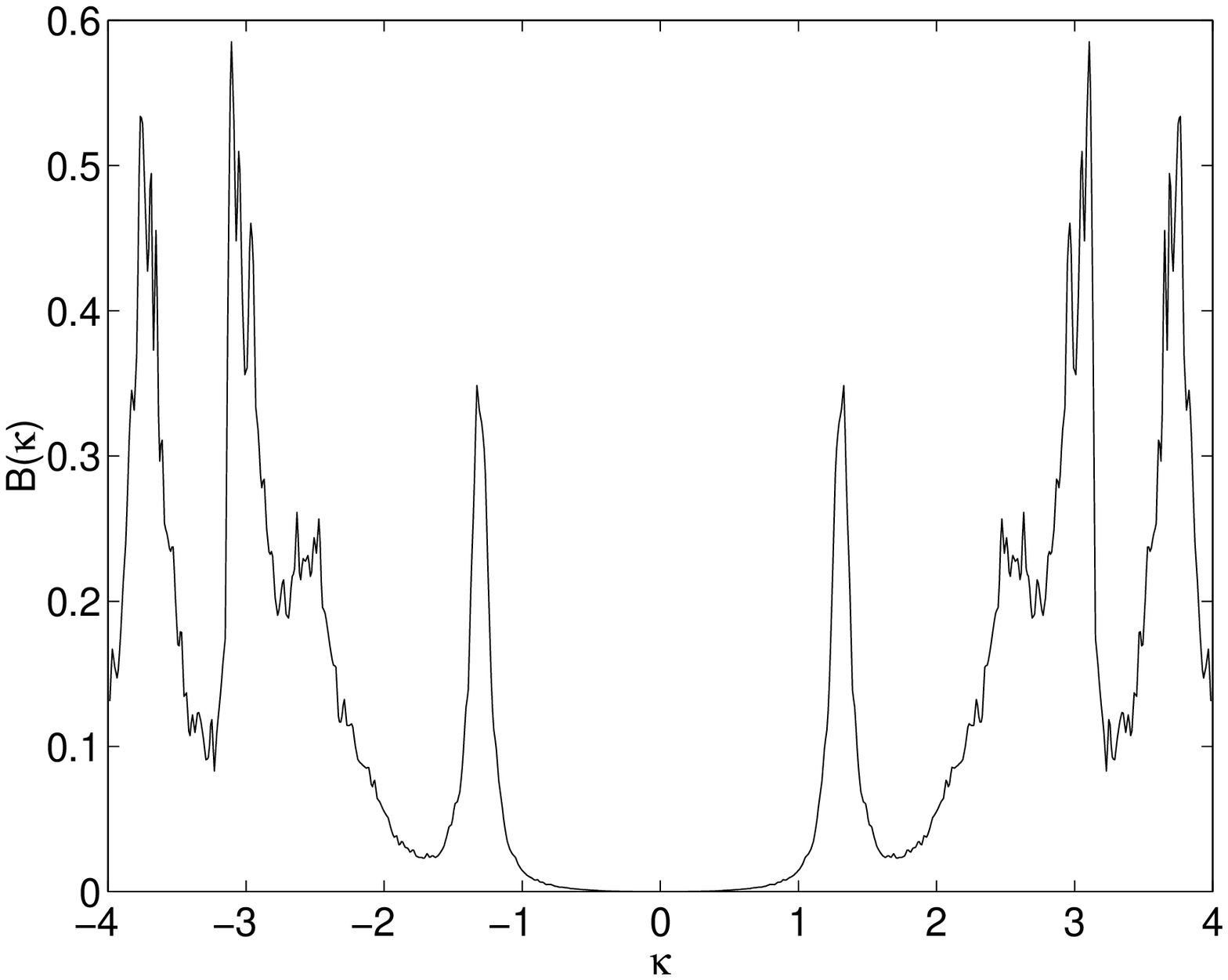,width=\hsize}}
\centerline{\epsfig{figure=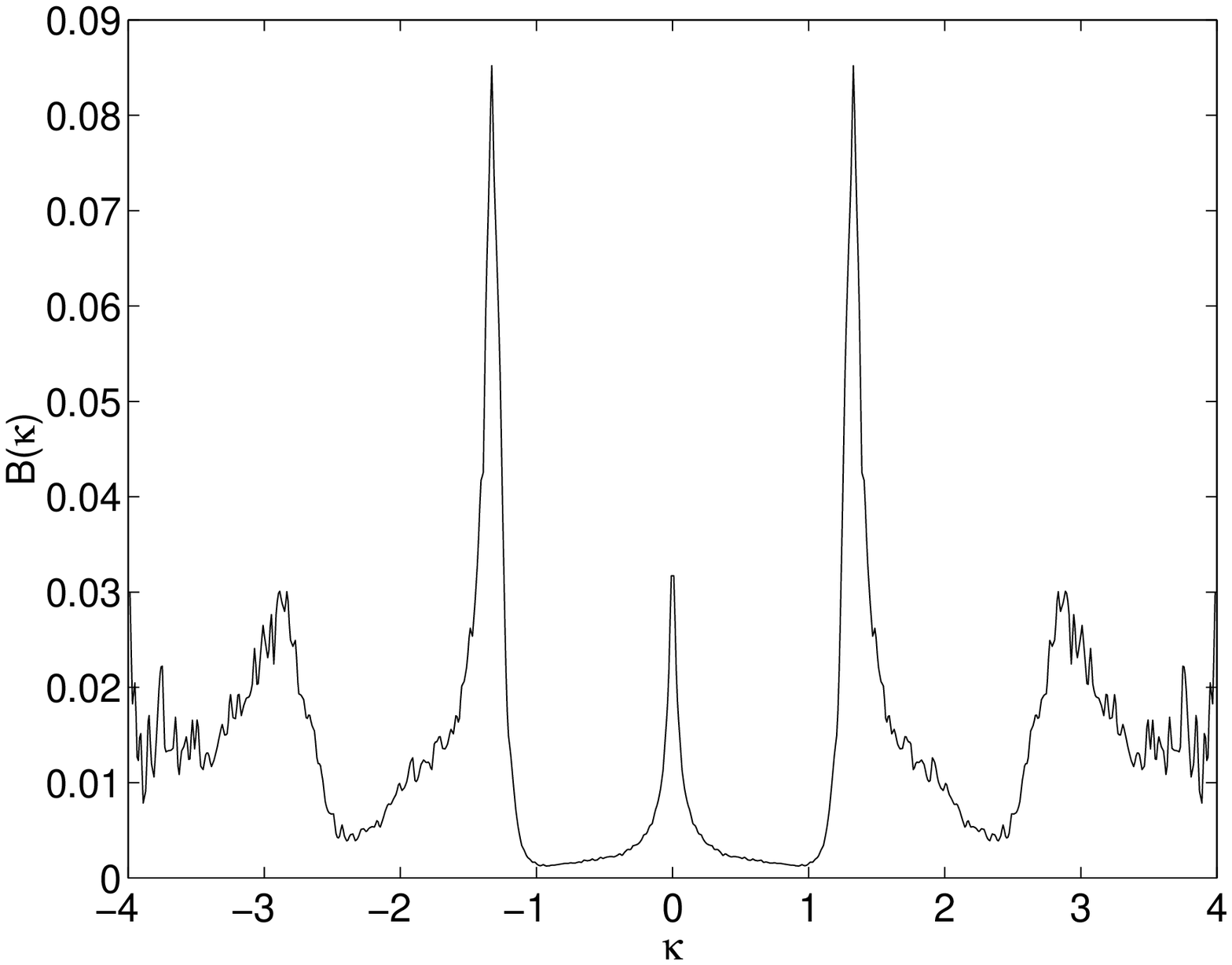,width=\hsize}}
\vspace{.1in}
\caption{
Band profiles for deformations of the quarter stadium 
as defined by Eq.(\ref{e16}).  
{\em Upper plot:} Rotation around the stadium center. 
{\em Lower plot:} Generic (non-special) deformation  
involving displacement of the curved edge. 
It is important to notice that for the special 
deformation we have $\tilde{C}(\omega)\rightarrow0$
in the limit $\omega\rightarrow0$. } 
\end{figure}

As a numerical example we have picked the 
stadium billiard. We have found the eigenstates 
of a de-symmetrized (quarter) stadium as described 
in Ref.\cite{dil}. We have selected those  
eigenstates whose eigen-energies $k_n$ are 
in the vicinity of $k=400$.   
Our two representative deformations are: 
{\bf (a)} rotation around the stadium center; 
and {\bf (b)} generic (non-special) deformation 
involving the curved edge. In the latter 
case the curved edge of the quarter stadium 
($0<s<\pi/2$) is pushed outwards with $D(s)=(\cos(s))^2$, 
while for the straight edges $D(s)=0$. 
(The corner $s=0$ is the $90^{\circ}$ intersection 
of the curved edge with the long straight edge). 
The respective band profiles are displayed in Fig.3.   
The band profile has been defined as  
\begin{eqnarray} \label{e16}  
B(\kappa) \ = \ \frac{1}{4k^2} \ 
\left\langle \left| 
\oint \varphi_n(\mbf{s})\varphi_m(\mbf{s}) 
\ D(\mbf{s}) d\mbf{s}
\right|^2\right\rangle  
\end{eqnarray}
where $\kappa=(k_n-k_m)$ is the distance from 
the diagonal. Note that $B(\kappa)$ is just 
a scaled version of the semiclassical $\tilde{C}(\omega)$ 
as implied by Eq.(\ref{e13}) with (\ref{e6}).  
The remarkable agreement of $B(\kappa)$ with the 
semiclassical calculation has been demonstrated 
in \cite{dil,wlf}.

It is important to realize that in the hard wall limit 
(which is assumed here) the matrix  
$({\partial {\cal H}}/{\partial x})_{nm}$ 
is not a banded matrix.  It would become banded 
if we were assuming {\em soft} walls. 
For soft walls $\tilde{C}(\omega)$ becomes vanishingly small 
for $\omega \gg v/\ell$.  The bandwidth in energy units is 
$\Delta_b=\hbar v/\ell$, and in dimensionless 
units it is 
\begin{eqnarray} \label{e17}  
b \ \ = \ \ \frac{\Delta_b}{\Delta} \ \ = \ \ 
\frac{{\mathsf V}}{\ell \ \lambda_{\tbox{B}}^{d{-}1}}
\end{eqnarray}
Unless stated otherwise we have $b=\infty$.

\section{Parametric evolution - numerical results}

The parametric evolution of $P(r)$ for rotation 
and for generic deformation of the stadium is illustrated 
by the images of Fig.4 and by the plots of Figs.5-6. 
The calculation of each $P(r)$ profile is carried 
out as follows: Given $\delta x$ we use the method 
which is described in \cite{alexthesis} in order to calculate 
the matrix $P(n|m)$. Then we plot the elements 
of $P(n|m)$  versus $\kappa = (k_n(x)-k_m(0))$. 
In order to obtain the average profile the plot 
is smeared using standard procedure (see remark \cite{note}). 
The transformation from $\kappa$ to $r=(n-m)$ is 
done using the relation (see remark \cite{note}): 
\begin{eqnarray} \label{e18}
\kappa \ \ = \ \ \tilde{\Delta} \cdot r \ - \ 
\frac{1}{d} k \times \frac{\delta {\mathsf V}}{{\mathsf V}} 
\end{eqnarray}
Above $\tilde{\Delta}$ is the mean level spacing 
of the $\{k_n\}$ spectrum, and $\delta {\mathsf V}$ 
is the volume change that is associated with 
the deformation (it is approximately proportional 
to $\delta x$). If the deformation is volume preserving 
(as in the case of rotation) then the second term 
equals zero. But for the generic deformation that 
we have picked in our second numerical example, 
the volume is not preserved, and the systematic 
`downwards' shift of the levels should be taken 
into account.

\begin{figure}
\centerline{\epsfig{figure=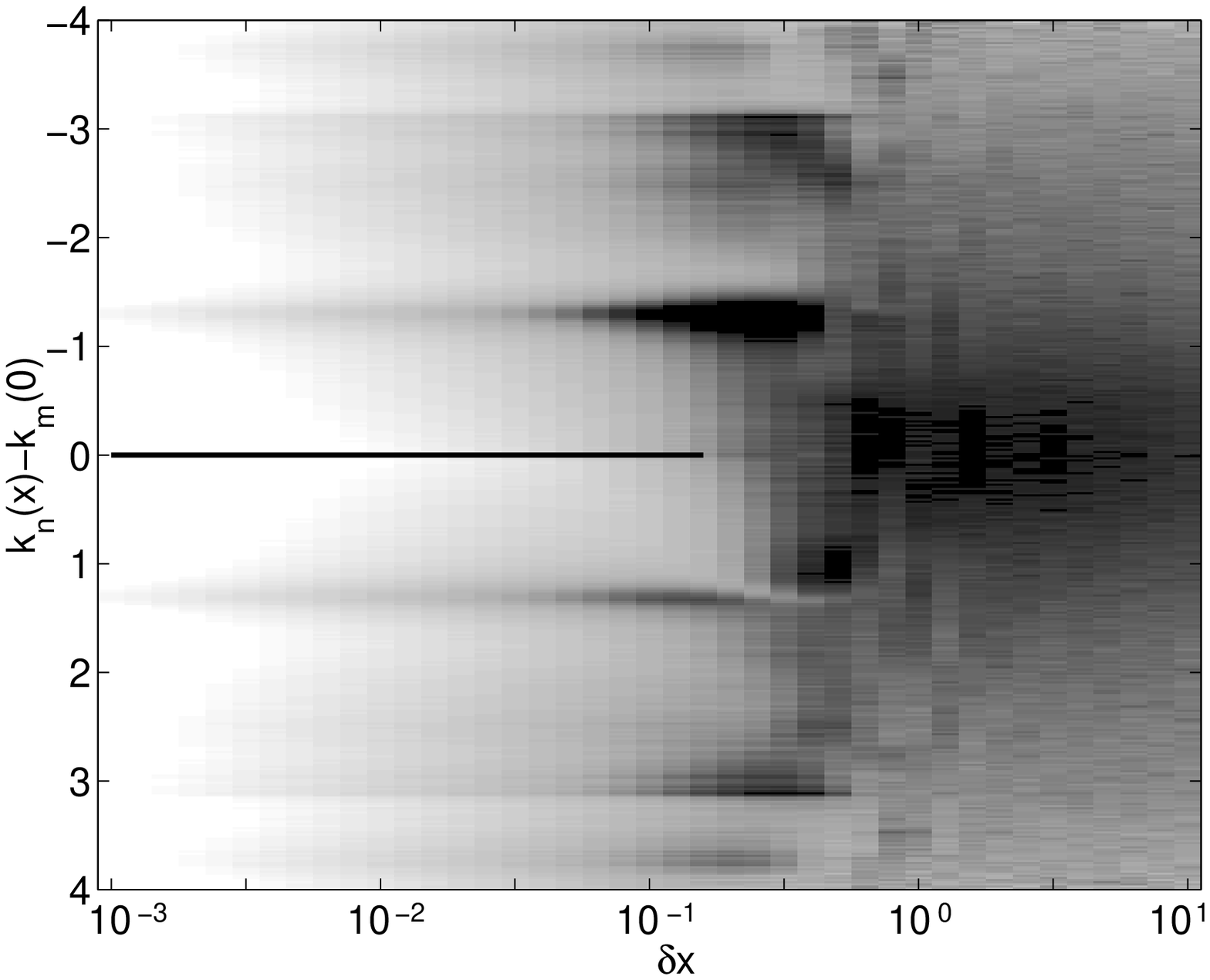,width=\hsize}}
\centerline{\epsfig{figure=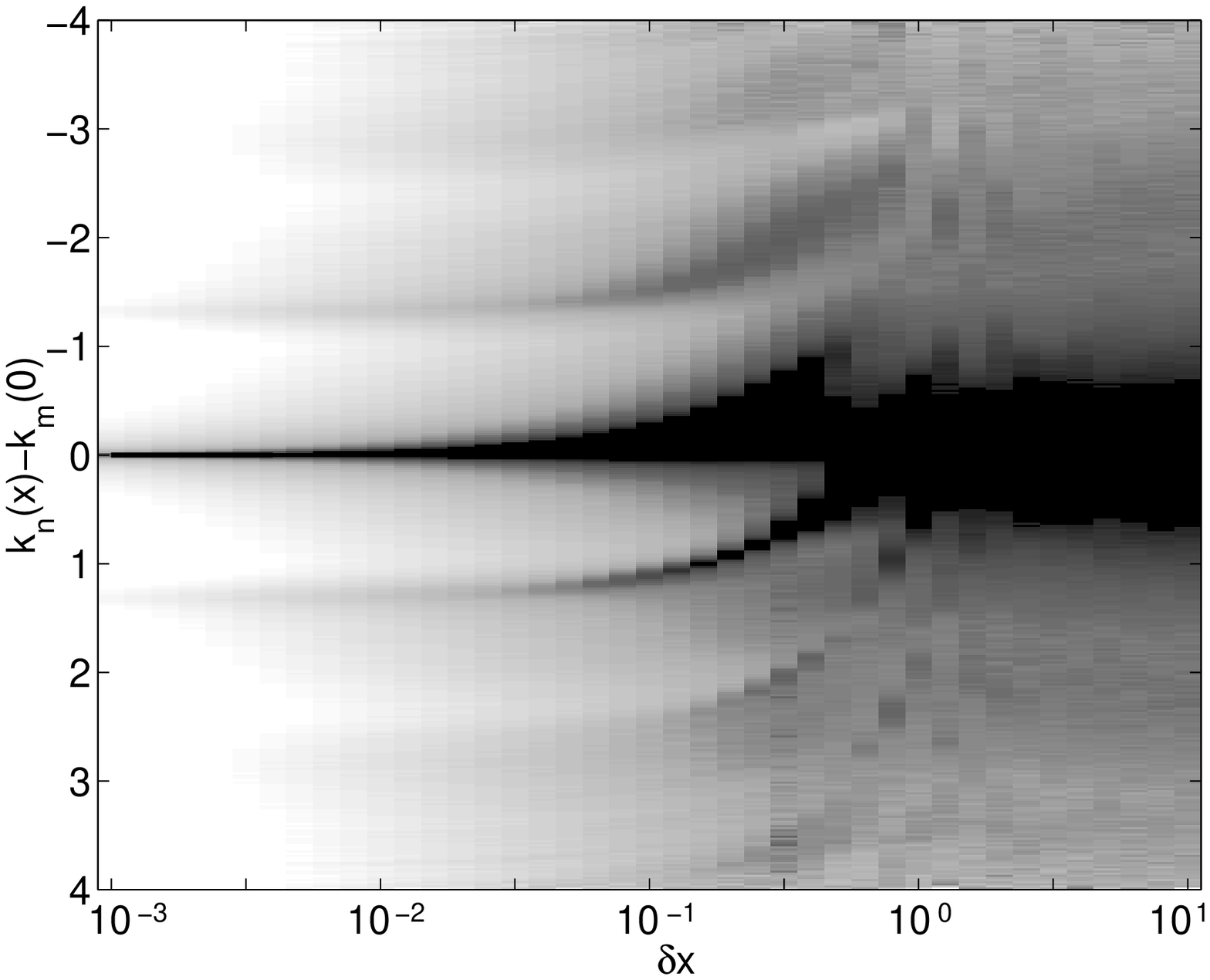,width=\hsize}}
\vspace{.1in}
\caption{
Each column is an image of $P(r)$ versus $\kappa$ 
for a different value of $\delta x$. There are 41 columns.  
The value $\delta x \sim 1$ corresponds roughly 
to $\lambda_{\tbox{B}}$ displacement of the boundary.  
The upper figure is for rotation, and the lower 
is for generic deformation. Note that the $r=0$ 
component is excluded from the image. Instead 
we have plotted over the image an $r=0$ line 
wherever this component contains more than $50\%$ 
of the probability. In the lower figure this line 
cannot be resolved from the developing core region.} 
\end{figure}

Looking first in the case of rotation,  
we see clearly two parametric regimes: 
The standard perturbative regime ($\delta x<0.2$),  
and the non-universal regime ($\delta x>0.2$). 
Let us clarify this observation. 
We see that for $\delta x<0.2$ most of the 
probability is well concentrated in $r=0$. 
This implies that we can use standard perturbation 
theory in order to estimate the $r \ne 0$ probabilities. 
On the other hand for $\delta x>0.2$ the perturbative 
nature of $P(r)$ is destroyed. Now $P(r)$ becomes smoother, 
and eventually (for $\delta x>0.5$) there is a very good 
fitting with Lorentzian (see lower plot in Fig.5).

The qualitative explanation for the 
Lorentzian profile is as follows: 
For $\delta x>0.5$ the typical displacement 
of the walls is of the order of $\lambda_{\tbox{B}}$. 
Therefore the $|n(x)\rangle$ eigenstates 
become uncorrelated with the $|m(0)\rangle$ 
eigenstates. Consequently $P(r)$ becomes 
$\delta x$ independent. The Lorentzian profile 
agrees with the assumption of uncorrelated 
random waves as explained in Appendix A.

Let us look now in the case of generic deformation. 
Here we see clearly three parametric regimes: 
The standard perturbative regime ($\delta x<0.004$), 
the core-tail regime ($0.004< \delta x < 0.2$), 
and the non-universal regime ($\delta x>0.2$). 
Let us clarify this observation.
As in the case of rotation there is a standard 
perturbative regime ($\delta x<0.004$) where most 
of the probability is well concentrated in $r=0$. 
For larger deformation, namely for $\delta x>0.004$, 
standard perturbation theory is no longer 
applicable because the $n=m$ level is mixed
non perturbatively with other (neighboring) levels. 
As a result $P(r)$ contains a non-perturbative 
`core' component. However, for $0.004< \delta x < 0.2$ 
we definitely do not get a Lorentzian. Rather the 
tails of $P(r)$ keep growing in the same way 
as in the standard perturbative regime.

In the case of the generic deformation, 
as in the case of rotation, 
we enter the non-universal regime, 
and eventually (for $\delta x>0.6$)
we get a smooth Lorentzian-like distribution. 
However, the Lorentzian-like distribution 
is not identical with that of the rotation case. 
Also the similarity to proper Lorentzian is  
far from being satisfactory. (see lower plot in Fig.6).
This means that the random-wave picture of Appendix A 
is an oversimplification.

In the following sections we are going to 
summarize the theoretical considerations \cite{wls} 
that explain the observed parametric scenario. 
In particular we are going to illuminate the way 
in which non-perturbative features emerge;  
to clarify the crossover to the non-universal 
regime; and to explain the specific nature of 
the non-universal distribution.

\begin{figure}[p]
\centerline{\epsfig{figure=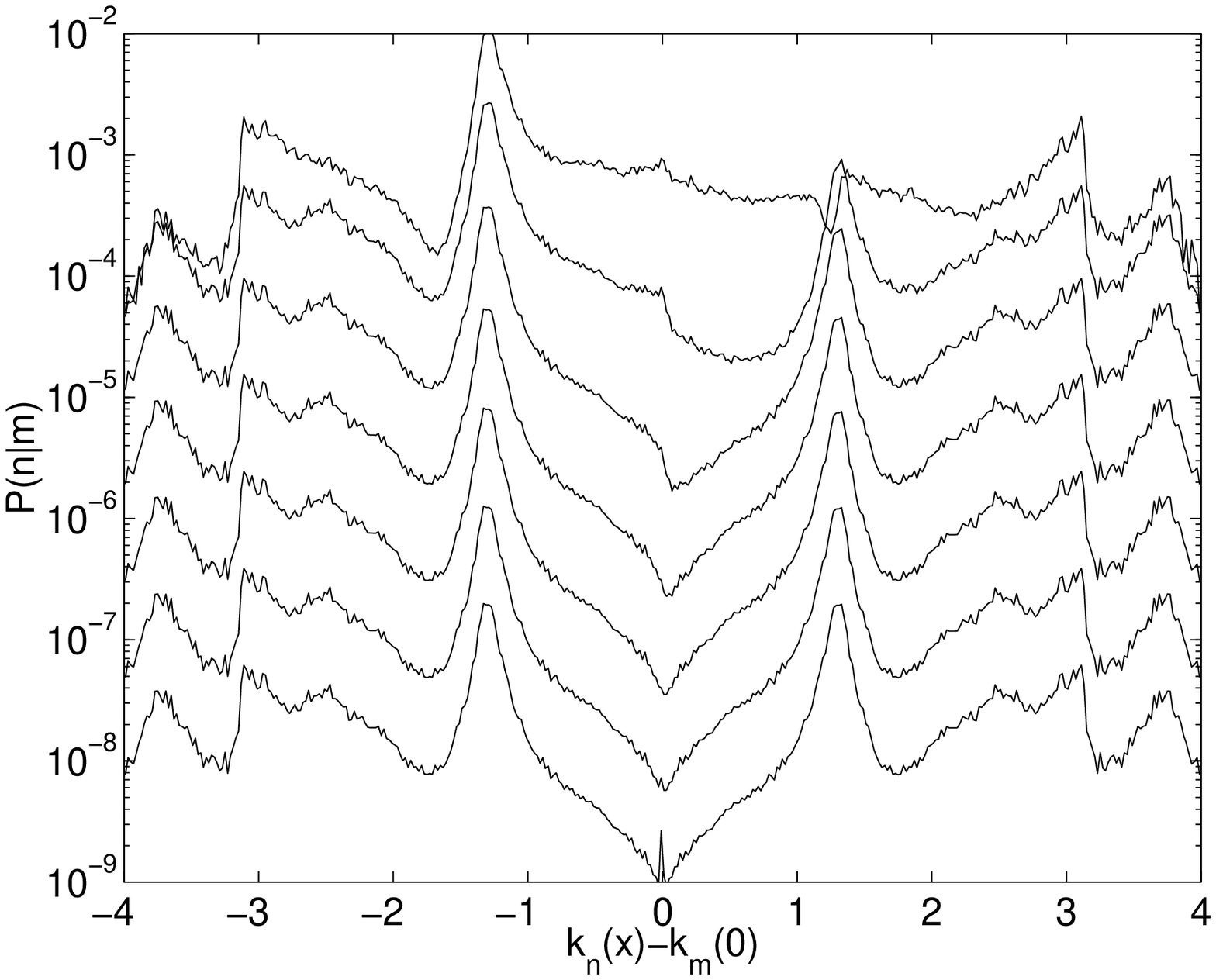,width=\hsize}}
\centerline{\epsfig{figure=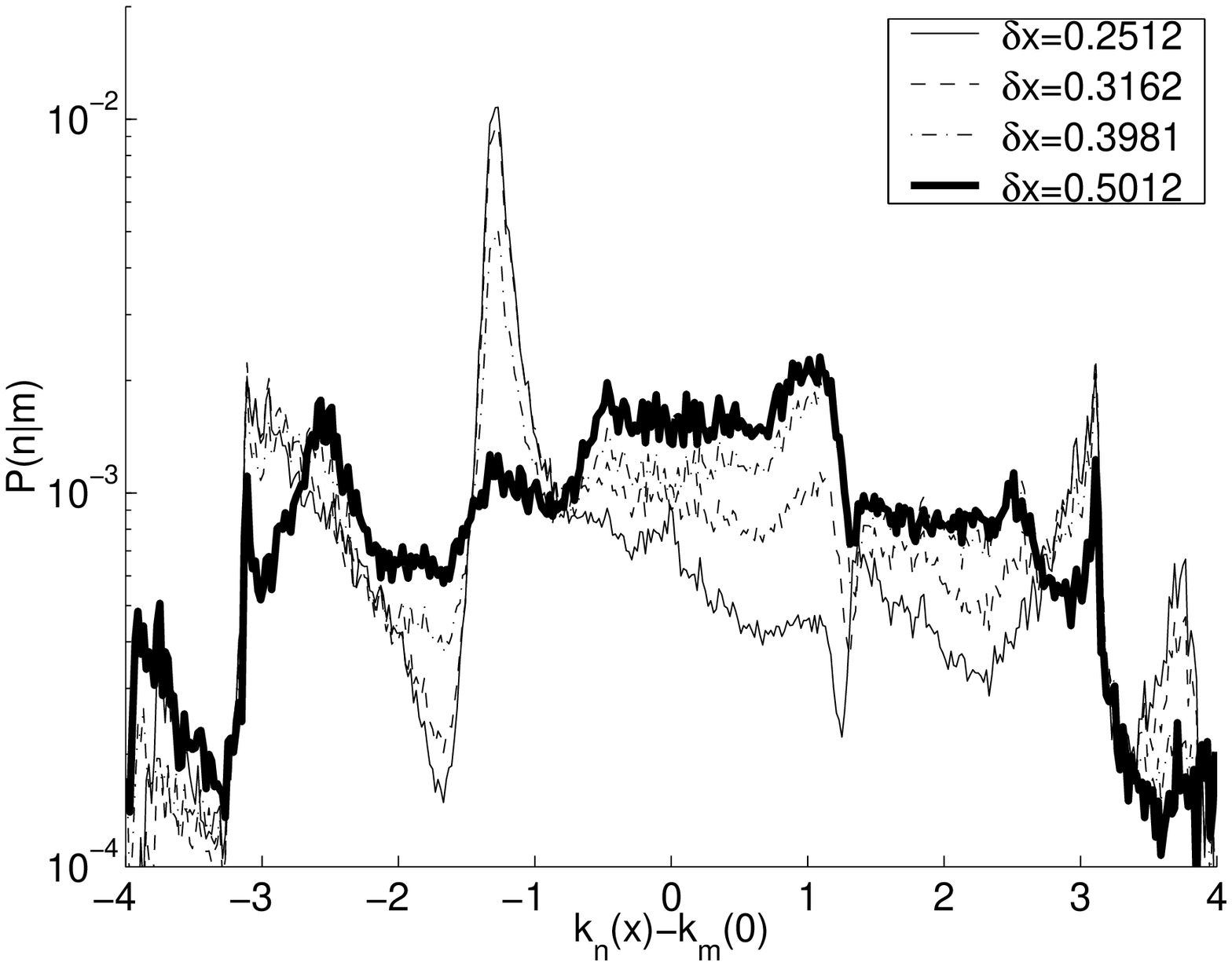,width=\hsize}}
\centerline{\epsfig{figure=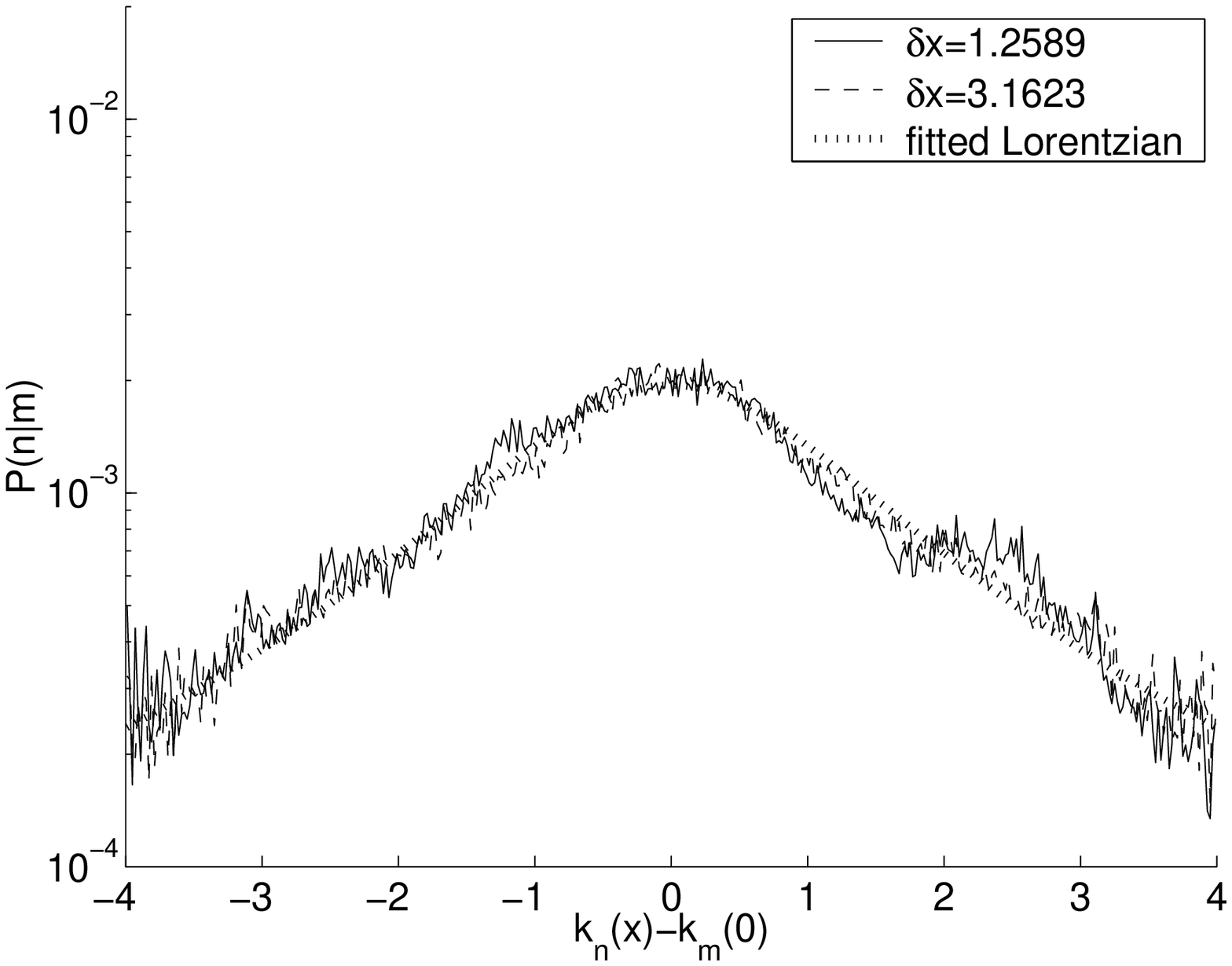,width=\hsize}}
\vspace{.1in}
\caption{
Representative plots of $P(r)$ for the case of 
rotation. The upper subfigure is for  
$0.0010 \le \delta x \le 0.2512$. The $r=0$ 
component is excluded. In the lower subfigure the 
fitted  Lorentzian is indistinguishable from the 
actual profile.}  
\end{figure}

\begin{figure}[p]
\centerline{\epsfig{figure=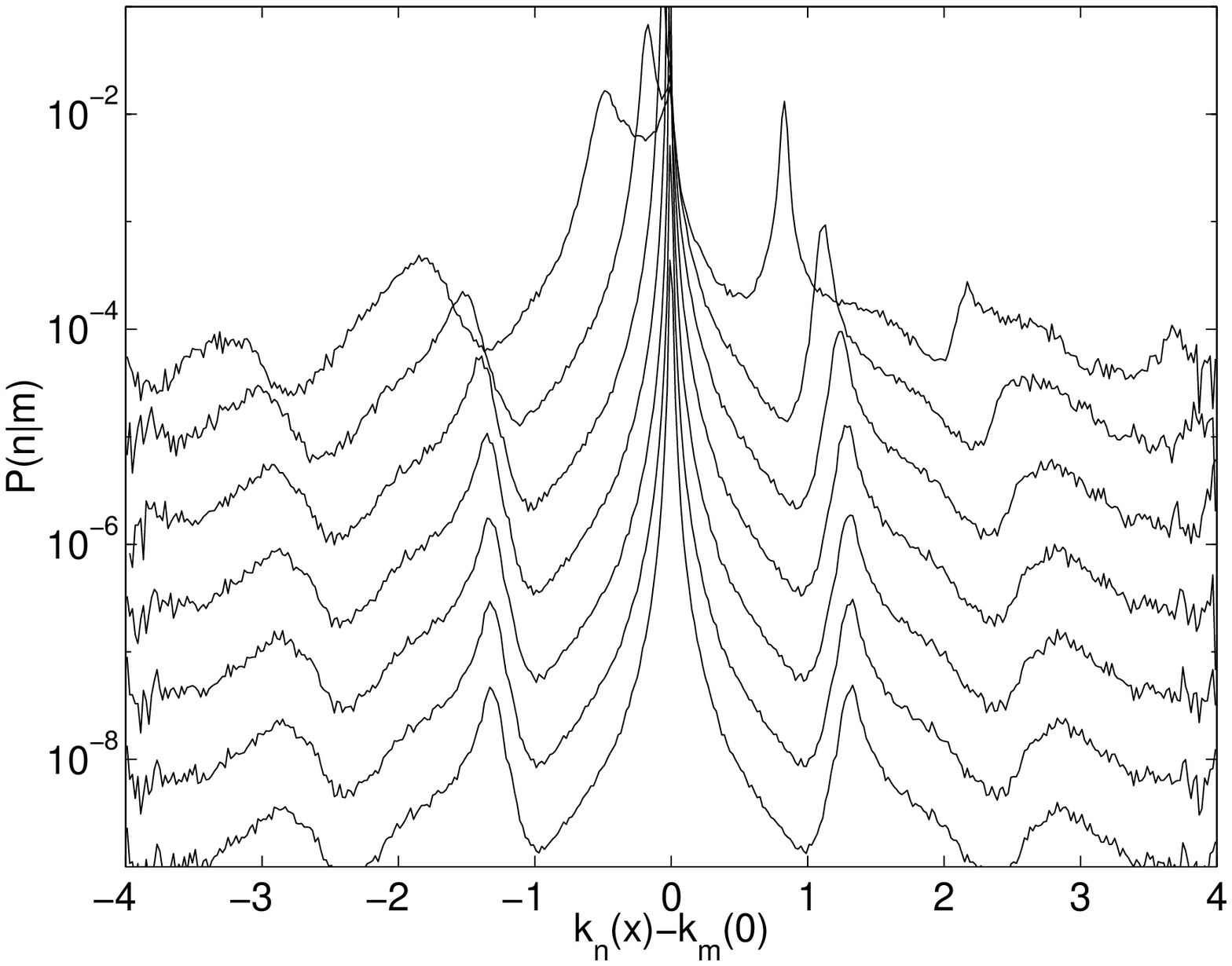,width=\hsize}}
\centerline{\epsfig{figure=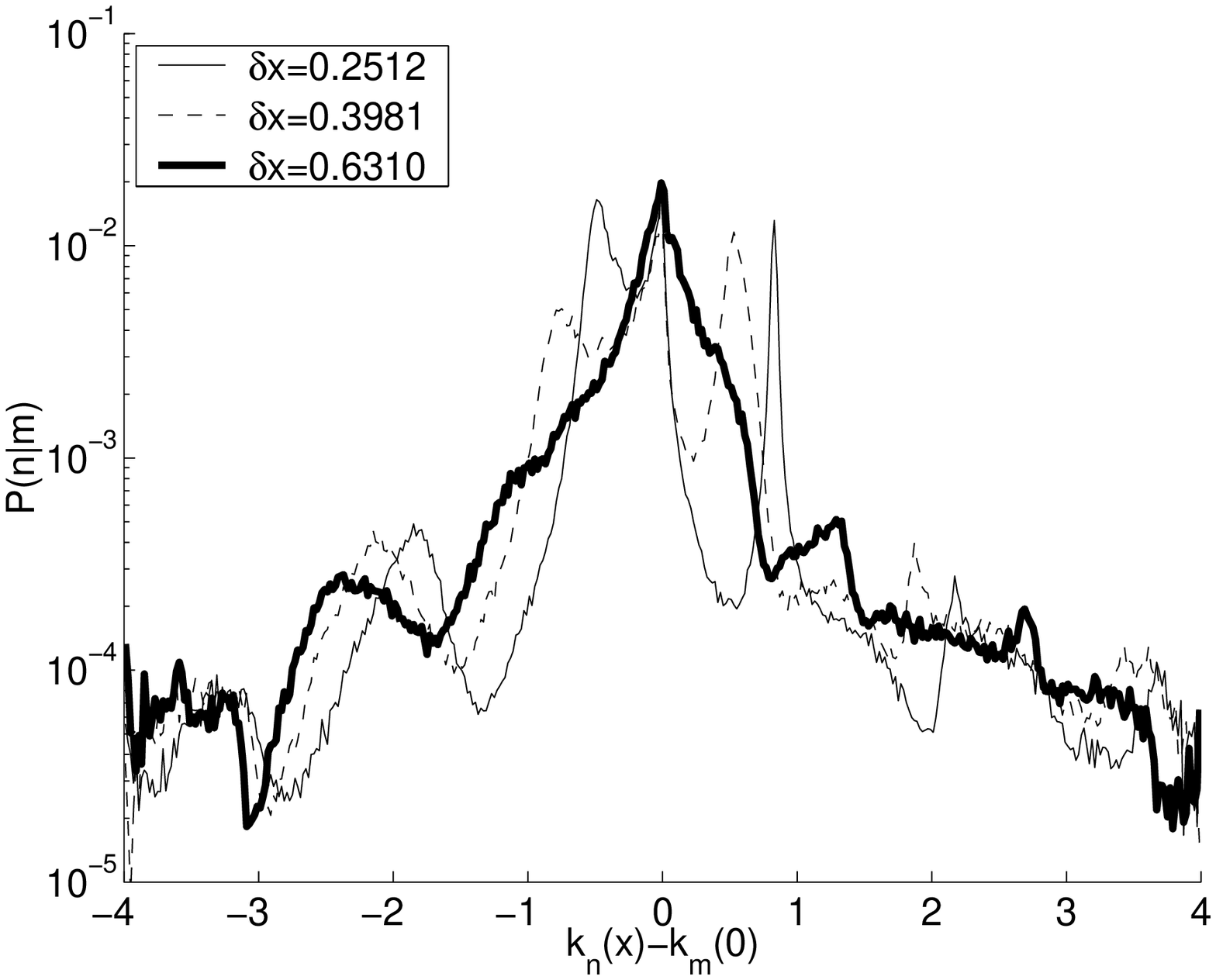,width=\hsize}}
\centerline{\epsfig{figure=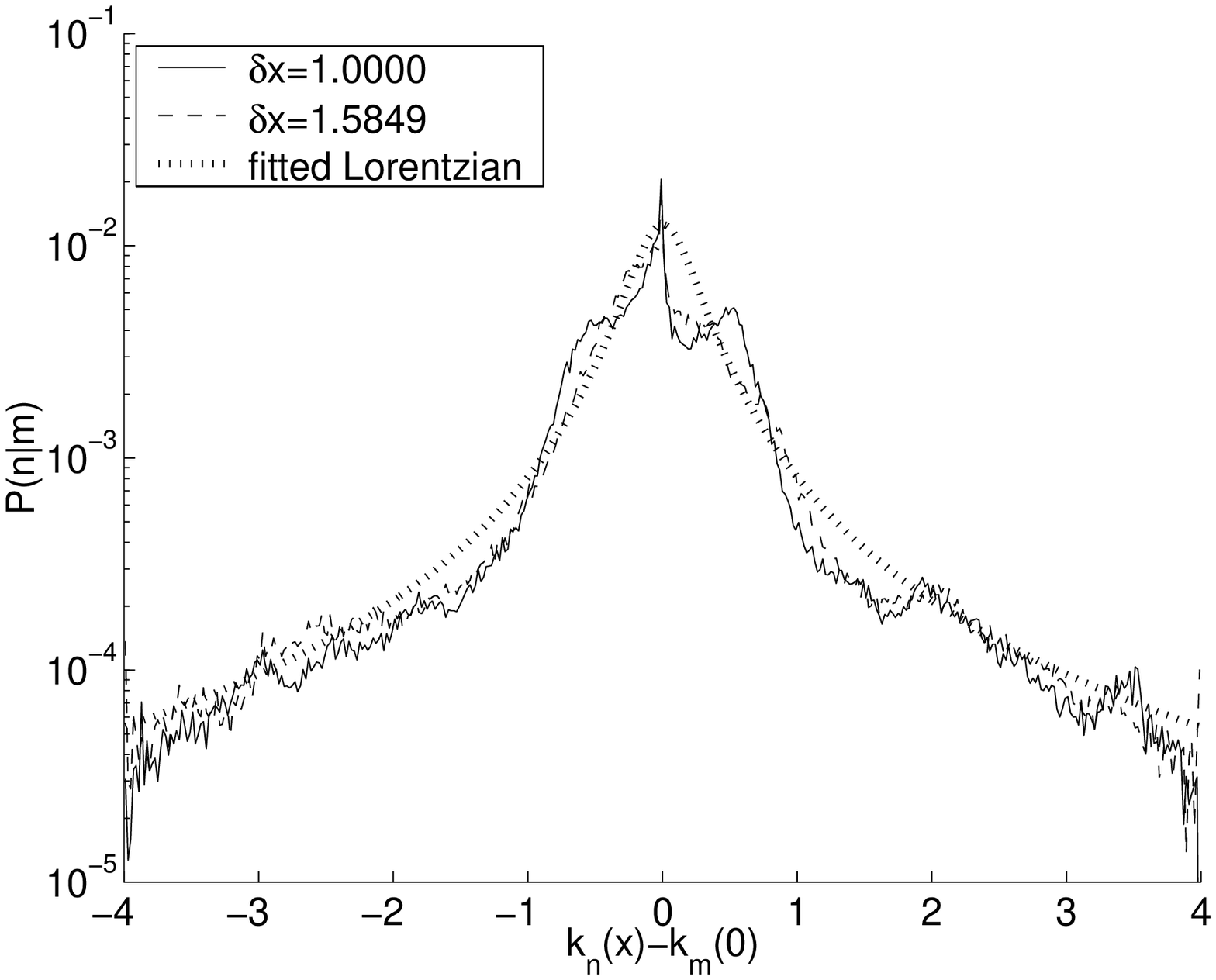,width=\hsize}}
\vspace{.1in}
\caption{
Representative plots of $P(r)$ for the case of 
generic deformation. The upper subfigure is for  
$0.0010 \le \delta x \le 0.2512$. In the lower 
subfigure a fitted Lorentzian is overlayed for 
the purpose of comparison.} 
\end{figure}

\section{The standard perturbative regime}     

Standard perturbation theory gives the following 
first order expression for the LDOS
\begin{eqnarray} \label{e19}
P(n|m) \ \ \approx \ \ \delta_{nm} \ + \ 
\delta x^2 
\frac{|({\partial {\cal H}}/{\partial x})_{nm}|^2}
{(E_n-E_m)^2}
\end{eqnarray}
This expression is most straightforwardly obtained 
by inspecting Eqs.(\ref{e3}) and (\ref{e4}). 
We can define the (total) transition probability as
\begin{eqnarray} \label{e20}   
p(\delta x) \ \ = \ \ \sum_{r\ne0} P(r)   
\end{eqnarray}
Using Eq.(\ref{e19}) combined with (\ref{e13}) 
we get the following estimate:
\begin{eqnarray} \label{e21}   
p(\delta x) \ \ \approx \ \ 
\delta x^2 \times \frac{1}{\hbar^2} 
\int_{|\omega|>\frac{\Delta}{\hbar}}\frac{d\omega}{2\pi}
\ \frac{\tilde{C}(\omega)}{\omega^2}    
\end{eqnarray}
Standard perturbation theory is applicable as 
long as $p(\delta x) \ll 1$. This can be converted 
into an equivalent inequality $\delta x \ll \delta x_c^{\tbox{qm}}$. 
By this definition $\delta x_c^{\tbox{qm}}$ is the 
parametric deformation which is needed in order 
to mix the initial level $m$ with other levels 
$n \ne m$.

If we use Eq.(\ref{e21}) for a special deformation,
then we have $\gamma>1$, and consequently 
the integral is not sensitive to the 
exclusion of the $|r|<\Delta$ region. 
As a result we have $\delta x_c^{\tbox{qm}} \propto \hbar$. 
Using Eq.(\ref{e15}) with (\ref{e14}) we get 
\begin{eqnarray} \label{e22} 
\delta x_c^{\tbox{qm}}|_{\mbox{special}} \ \ = \ \   
\left(\ell_{\tbox{bl}}\frac{1}{{\mathsf V}}
\oint[D(\mbf{s})]^2 d\mbf{s} 
\right)^{-\half} \times \lambda_{\tbox{B}}
\end{eqnarray}
In case of generic deformation Eq.(\ref{e22}) is not 
valid because the value of the integral 
in (\ref{e21}) is predominantly determined 
by the $\omega \sim \Delta/\hbar$ lower cutoff rather 
than by an effective $\omega \sim 1/\tau_{\tbox{bl}}$ 
lower cutoff. As a result one obtains 
\begin{eqnarray} \label{e23} 
\delta x_c^{\tbox{qm}}|_{\mbox{generic}} \ \ = \ \  
\left(\ell_{\tbox{H}}\frac{1}{{\mathsf V}}
\oint[D(\mbf{s})]^2 d\mbf{s} 
\right)^{-\half} \times \lambda_{\tbox{B}}
\end{eqnarray}
where $\ell_{\tbox{H}}=vt_{\tbox{H}}$ is the length which is 
associated with the Heisenberg time
$t_{\tbox{H}}=2\pi\hbar/\Delta$.

It is illuminating to use the convention $D(\mbf{s})\sim 1$, 
such that $\delta x$ measures the typical 
displacement of a wall element. With this convention 
we get from Eqs.(\ref{e22})-(\ref{e23}) the following: 
\begin{eqnarray}
\label{e24_s}  
\delta x_c^{\tbox{qm}} & 
|_{\mbox{special \hspace*{3.5mm}}} \ \ & \approx  \ \ 
\lambda_{\tbox{B}}
\\ \label{e24} 
\delta x_c^{\tbox{qm}} & 
|_{\mbox{generic \hspace*{3.5mm}}} \ \ & =  \ \  
\left(
\frac{\lambda_{\tbox{B}}^{d{-}1}}{{\mathsf A}_{\tbox{eff}}}
\right)^{\half}
\times \lambda_{\tbox{B}}
\\ \label{e24_d}
\delta x_c^{\tbox{qm}} & 
|_{\mbox{diffractive}} \ \ & \sim  \ \ 
\lambda_{\tbox{B}}
\end{eqnarray}
In the generic case the effective area of the deformed 
boundary ${\mathsf A}_{\tbox{eff}}$ may be smaller than 
the total area ${\mathsf A}$ of the boundary. The 
effective area ${\mathsf A}_{\tbox{eff}}$ can be 
formally defined by comparing Eq.(\ref{e24}) with Eq.(\ref{e23}).
Eq.(\ref{e24_d}) has been added for sake of completeness of 
our presentation. It correspond to the diffractive limit 
${\mathsf A}_{\tbox{eff}} \rightarrow \lambda_{\tbox{B}}^{d{-}1}$. 
Note that this limit is beyond the scope of the present study.   
Thus in the generic case, the wall displacement needed to mix levels 
is much smaller than $\lambda_{\tbox{B}}$. 
In the generic case $\delta x_c^{\tbox{qm}} \propto \hbar^{(d{+}1)/2}$ 
rather than $\delta x_c^{\tbox{qm}} \propto \hbar$. 
What happens with perturbation theory beyond $\delta x_c^{\tbox{qm}}$? 
This is the subject of the next section.

\section{The core-tail regime}    

For $\delta x > \delta x_c^{\tbox{qm}}$ standard perturbation 
theory diverges due to the non-perturbative 
mixing of neighboring levels on small scale:  
Once $\delta x$ becomes of the order of $\delta x_c^{\tbox{qm}}$ 
several levels are mixed, and as $\delta x$ becomes larger, 
more levels are being mixed non-perturbatively. 
Consequently it is natural to distinguish between 
{\em core} and {\em tail} regions \cite{vrn,frc,lds}.
Most of the spreading probability is contained within
the core region, which implies a natural extension
of first-order perturbation theory (FOPT): 
The first step is to transform Eq.(\ref{e3})
into a new basis where transitions within the core are
eliminated; The second step is to use FOPT (in the new basis)
in order to analyze the core-to-tail transitions.
Details of this procedure, which is in the spirit of 
degenerate perturbation theory, were discussed in \cite{frc}. 
The most important (and non-trivial) consequence of this
procedure is the observation that mixing on small scales
does not affect the transitions on large-scales. 
Therefore we have in the tail region $P(n|m)\propto \delta x^2$
rather than $P(n|m)\propto \delta x$. 
The validity of this observation has been numerically  
illustrated in \cite{lds}.

Following the above reasoning we define 
the {\em tail} region as consists of those levels 
whose `occupation' can be calculated using perturbation 
theory, while the {\em core} is the non-perturbative 
component in the vicinity of $r=0$. 
Assuming that only one scale characterize the core width, 
one arrives at the following practical approximation:
\begin{eqnarray} \label{e25} 
P_{\tbox{prt}}(r) =  
\frac{\Delta}{2\pi\hbar} \ 
\tilde{C}\left(\frac{E_n{-}E_m}{\hbar}\right) \
\frac{\delta x^2}
{(\Gamma(\delta x))^2 + (E_n{-}E_m)^2}
\end{eqnarray}
It is implicit in this definition that $(E_n{-}E_m)$ 
should be regarded as a function of $r=(n-m)$. 
The parameter $\Gamma(\delta x)$ is determined 
(for a given $\delta x$) such that $P_{\tbox{prt}}(r)$ 
has a unit normalization. One may say that 
$\Gamma(\delta x)$ regularizes the behavior 
around $r=0$. For generic deformation we get
\begin{eqnarray} \label{e26} 
\Gamma(\delta x) \ \equiv \ b_0\Delta \ \approx \ 
\left(\frac{\delta x}{\delta x_c^{\tbox{qm}}}\right)^2 \times \Delta 
\end{eqnarray}
The core is defined as the region $|r|<b_0$, 
and the outer ($|r|>b_0$) regions are the tails. 
For $\delta x \ll \delta x_c^{\tbox{qm}}$ we 
get $b_0 \ll 1$ and the core-tail structure Eq.(\ref{e25}) 
becomes equivalent to the standard 
perturbative result Eq.(\ref{e19}). 
It should be clear that the core-tail structure 
is a generalization of Wigner's Lorentzian. 
It is indeed a Lorentzian in the 
special case of a `flat' band profile.

\begin{figure}
\centerline{\epsfig{figure=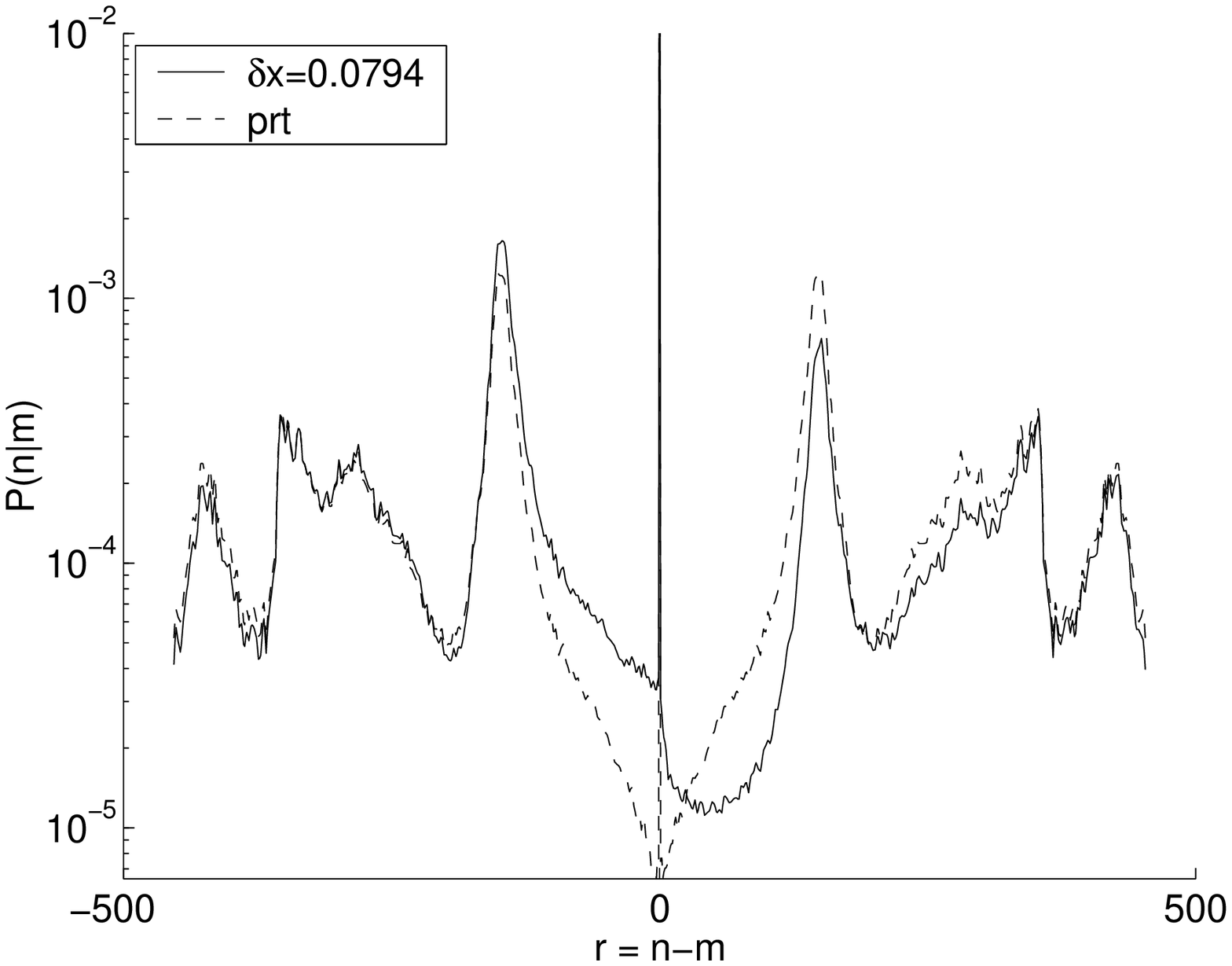,width=\hsize}}
\centerline{\epsfig{figure=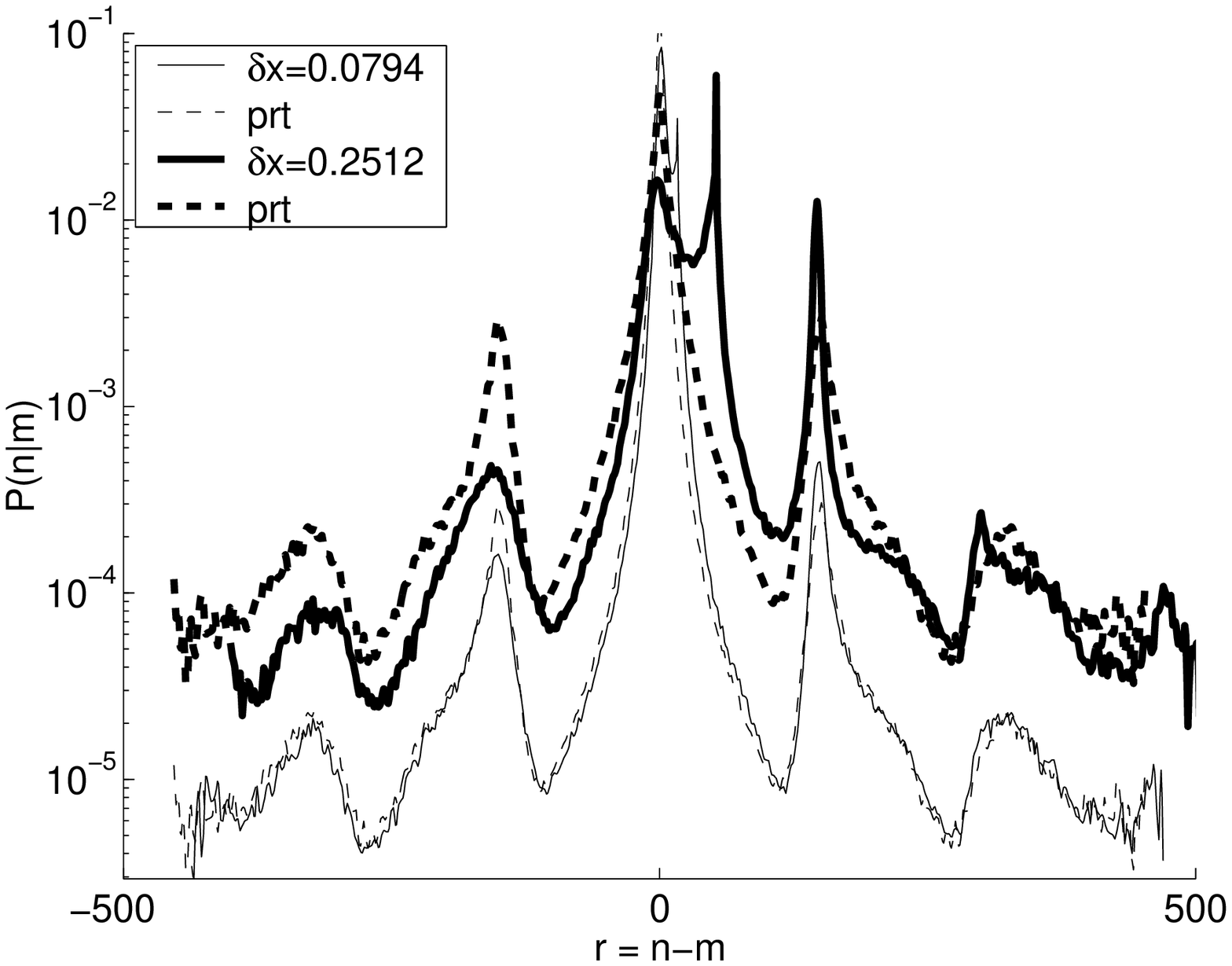,width=\hsize}}
\centerline{\epsfig{figure=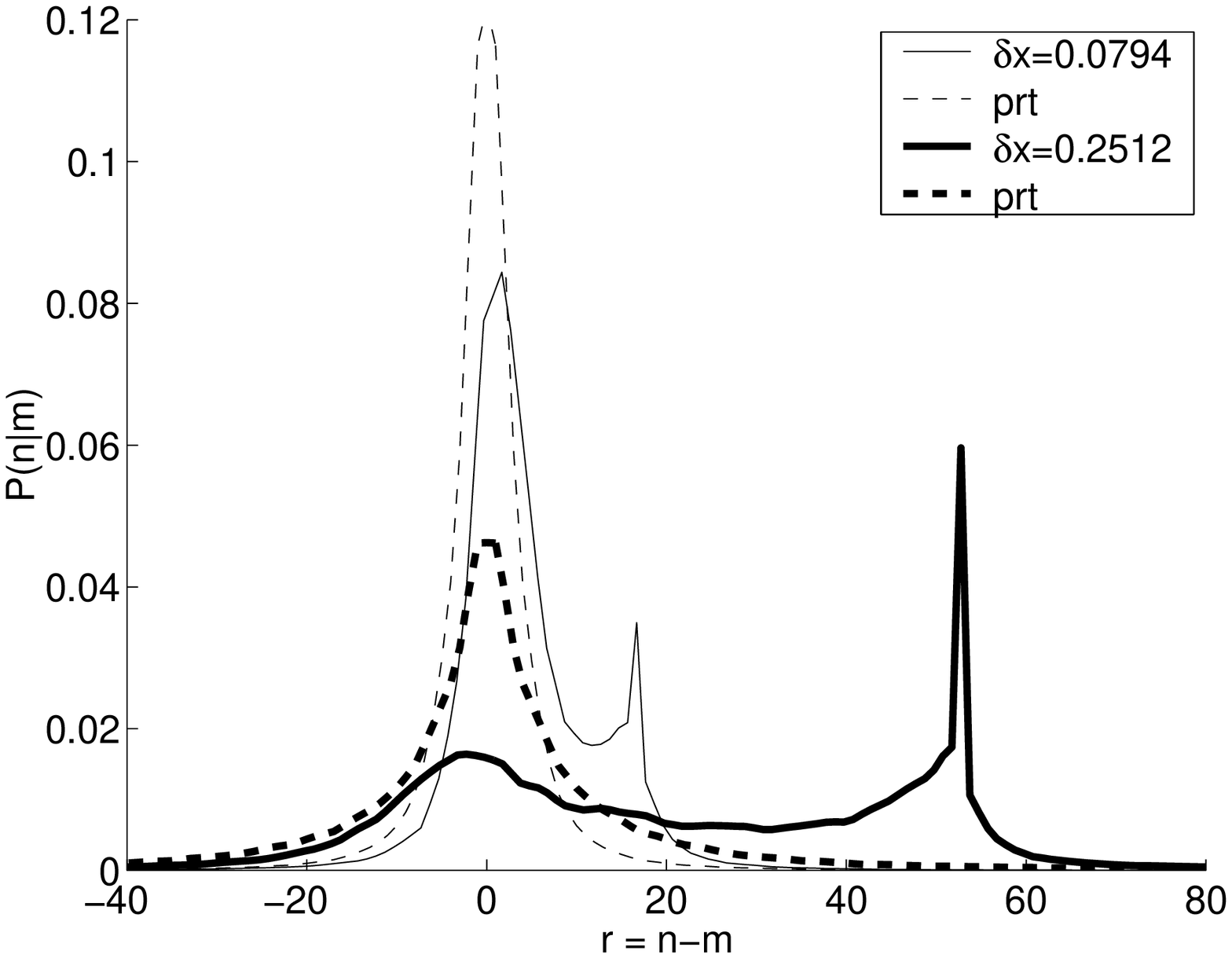,width=\hsize}}
\vspace{.1in}
\caption{
Comparison between the perturbative calculation  
$P_{\tbox{prt}}(r)$ and the actual $P(r)$. The upper 
subfigure is for the rotation and the lower subfigures 
are for the generic deformation. The lower subfigure 
gives a zoom over the the `birth' of the 
non-perturbative semiclassical saturation profile.} 
\end{figure}

\begin{figure}
\centerline{\epsfig{figure=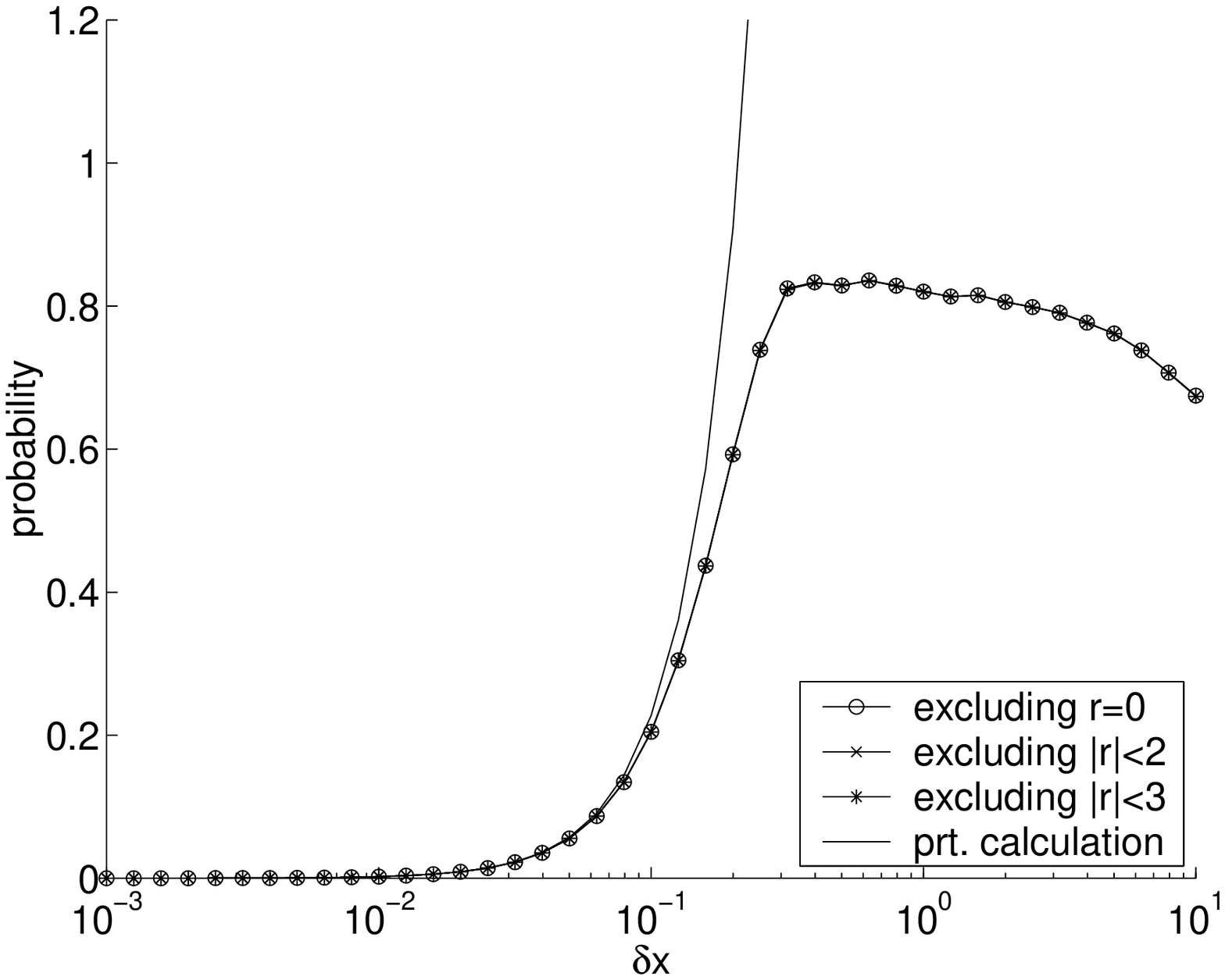,width=\hsize}}
\centerline{\epsfig{figure=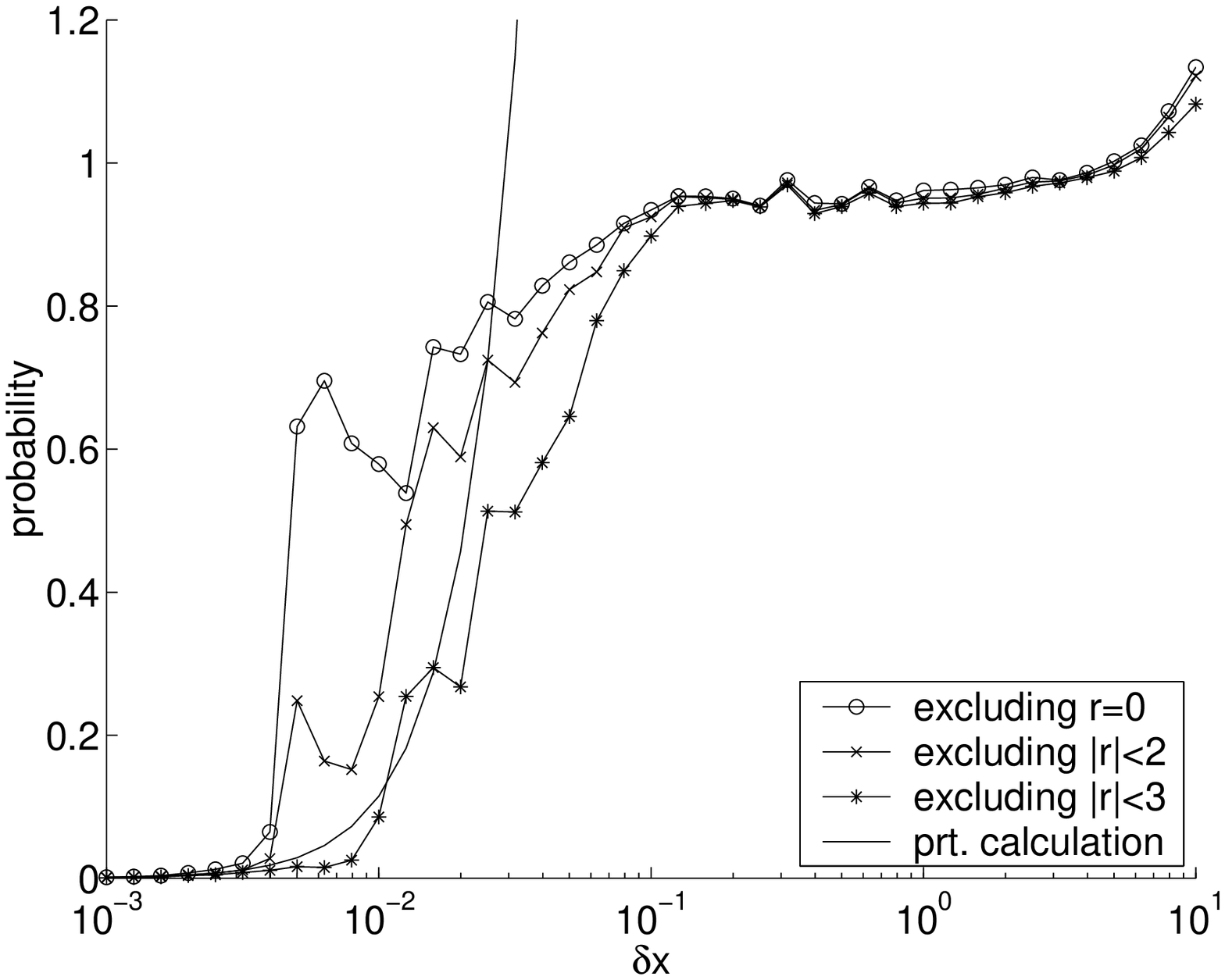,width=\hsize}}
\centerline{\epsfig{figure=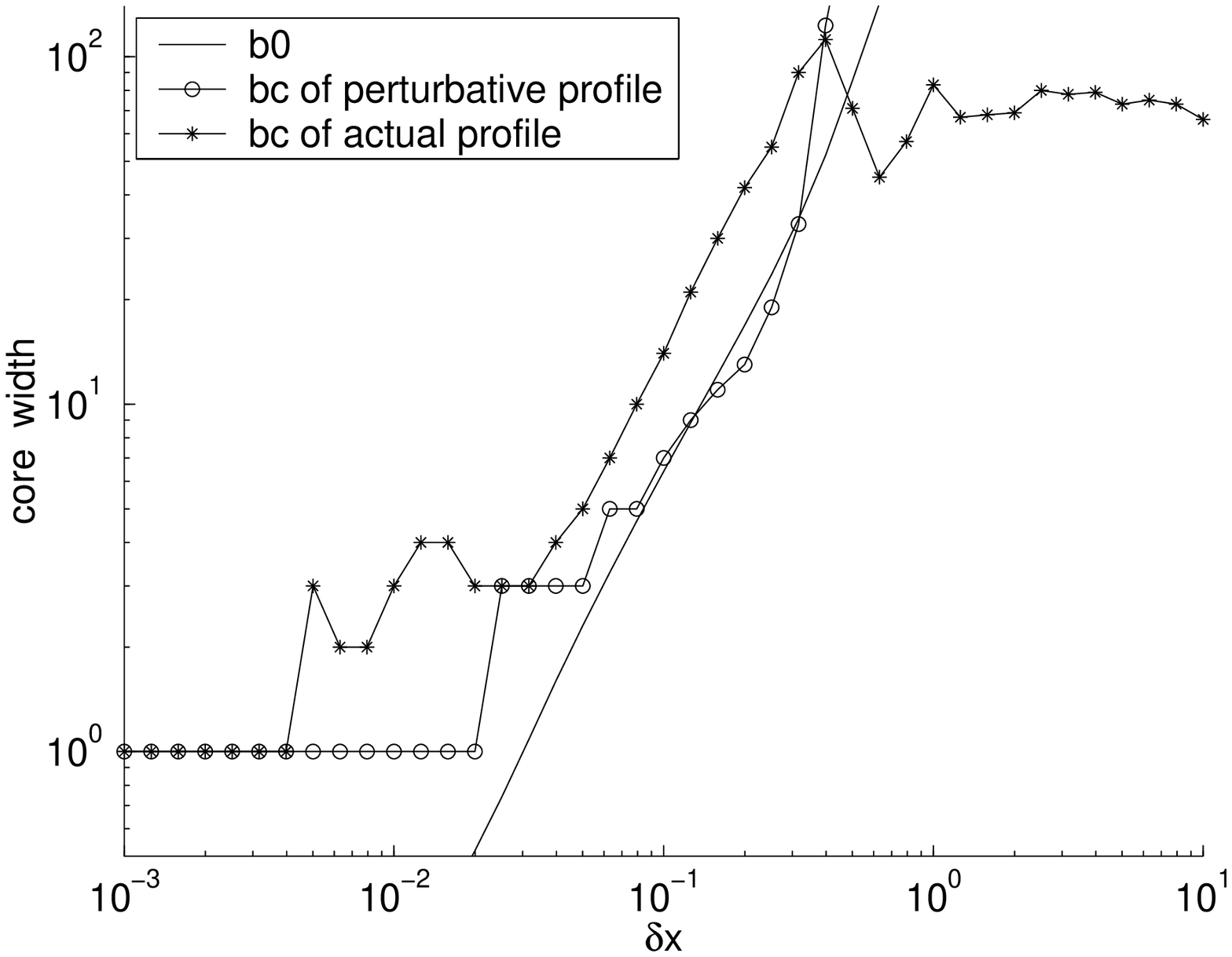,width=\hsize}}
\vspace{.1in}
\caption{
The upper and the middle subfigures are $p(\delta x)$ for 
the rotation and for the generic deformation respectively.  
The lower subfigure is the perturbative calculation  
of the core width $b_0$ for the generic deformation. 
Also displayed is the calculated $b_c$ for $P_{\tbox{prt}}(r)$,  
and the actual $b_c$ for $P(r)$.  } 
\end{figure}

We turn now to analyze our numerical results. 
We have verified (see eg Fig.7) 
that for $\delta x < 0.2$ we have   
good agreement with perturbation theory 
irrespective of whether we have a core component 
(which is the case for the generic deformation) 
or not (which is the case for the rotation). 
As we come closer to $\delta x \sim 0.2$ 
the agreement becomes worse, and for $\delta x > 0.3$ 
we have a total collapse of perturbation theory.
In Fig.8 we display the total transition 
probability $p(\delta x)$ as a function 
of $\delta x$. This plot should be used in order 
to numerically determine the value of $\delta x_c^{\tbox{qm}}$, 
say as the value where $p(\delta x)=1/2$.

Fig.8 also displays comparison with the corresponding 
perturbative calculation 
(using Eq.(\ref{e20}) with(\ref{e25})). 
The agreement (for $\delta x < \delta x_c^{\tbox{qm}}$) 
in case of the rotation is remarkable.
For larger values of $\delta x$ one observes  
a linear drop of the total probability due to the non-unitary 
nature of the evolution. (This drop is clearly linear 
in a linear-linear scale which is not displayed).   
The eventual rise of the total transition probability 
(beyond $1$) in case of the generic deformation, 
reflects numerical errors in the determination 
of the small $\kappa$ overlaps. 
See \cite{alexthesis} for further details.

The lower subfigure in Fig.8 displays the 
calculated core width $b_0$ for the 
generic deformation. Recall that $b_0$ is determined, 
given $\delta x$, such that $P_{\tbox{prt}}(r)$ of Eq.(\ref{e25}) 
is normalized. Also calculated is the width $b_c$ 
of the region that contains $50\%$ of the probability. 
The width $b_c$ is calculated for both $P_{\tbox{prt}}(r)$
and $P(r)$. Note that $b_c= 1$ in the standard perturbative 
regime.  The determination of $\delta x_c^{\tbox{qm}}$ for the 
generic deformation becomes more convenient by  
using this plot. Also the crossover 
(at $\delta x_{\tbox{NU}} \sim 0.2$) 
to the non-universal regime is most pronounced.

\section{The non-universal regime}

The validity of Eq.(\ref{e25}) as a global approximation 
relies on the assumption that the core is 
characterized only by one scale (which is $b_0$). 
But this assumption ceases to be true if 
$\delta x$ is large enough. We have explained in 
Ref.\cite{wls} that the width of the {\em core}
defines a `window' through which we can view
the `landscape' of the semiclassical analysis.
As $\delta x$  becomes larger, this `window' becomes wider, 
and eventually some of semiclassical structure is exposed. 
This is marked by the non-universal parametric scale $\delta x_{\tbox{NU}}$.
For $\delta x$ larger than $\delta x_{\tbox{NU}}$,   
the non-universal structure of the core is exposed.

What is the semiclassical structure of the LDOS? 
Time-domain semiclassical considerations 
(See Eq.(10) of \cite{wls} and related discussion there) 
imply that the non-universal structure of the core is 
\begin{eqnarray} \label{e27}
P(n-m) \ \ \approx \ \ \frac{1}{\pi} \
\frac{\delta E_{\tbox{SC}}}{\delta E_{\tbox{SC}}^2+(E_n{-}E_m)^2}
\end{eqnarray}
with $\delta E_{\tbox{SC}}=\hbar/\tau_{\tbox{col}}$. 
The definition of the collision rate $1/\tau_{\tbox{col}}$ 
is similar to that of $1/\tau_{\tbox{bl}}$. 
The former is the collision rate with the deformed area  
${\mathsf A}_{\tbox{dfr}}$, and therefore it may be 
smaller than  $1/\tau_{\tbox{bl}}$, because $\tau_{\tbox{bl}}$
is related to the total area ${\mathsf A}$.

If, by mistake, we identified $\tau_{\tbox{col}}$
with $\tau_{\tbox{bl}}$, then we would get the 
random wave result which is derived in Appendix A. 
This would be an over simplification.  If it were true, 
it would imply that for $\delta x > \delta x_{\tbox{NU}}$
we should have got the same  
Lorentzian-distribution in both cases 
(the rotation and the generic deformation). 
What we see, as a matter of fact, is that for the rotation 
(Fig.5) we have a reasonably good agreement with Lorentzian 
whose width is $\delta k_{\tbox{SC}}=1.5$, whereas 
for the generic deformation (Fig.6) there is rough 
agreement with Lorentzian whose width is $\delta k_{\tbox{SC}}=0.26$. 
The smaller width in the latter case clearly reflects 
having larger $\tau_{\tbox{col}}$. 
For the generic deformation we also have pronounced 
non-Lorentzian features. Actually the global fitting 
to Lorentzian is quite bad. Our understanding 
is that these features are due to the bouncing-ball trapping: 
It leads to non-exponential decay of the time-dependent 
survival probability (see \cite{wls} for definition 
of the latter term), and hence to the observed 
non Lorentzian features of the spreading profile.

We turn now to explain how  $\delta x_{\tbox{NU}}$ 
is determined. By definition it is the 
deformation which is required in order to expose 
features of the semiclassical landscape. 
These features start to be exposed once 
$\Gamma(\delta x) \sim \delta E_{\tbox{SC}}$ which 
should be converted into an equivalent 
expression $\delta x \sim \delta x_{\tbox{NU}}$.  
Thus we get 
\begin{eqnarray} \label{e28}
\delta x_{\tbox{NU}} \ \ = \ \ 
\left( 
\frac{{\mathsf A}_{\tbox{dfr}}}{{\mathsf A}_{\tbox{eff}}} 
\right)^{\half}
\times \lambda_{\tbox{B}} \ \ \sim \ \  \lambda_{\tbox{B}}
\end{eqnarray}
Here  ${\mathsf A}_{\tbox{dfr}}$ is the geometric area of the 
deformation (in the sense of scattering cross-section), 
while ${\mathsf A}_{\tbox{eff}}$ is the effective area of the 
deformation. The definition of the latter is implied 
by comparing Eq.(\ref{e24}) with Eq.(\ref{e23}).  
By rescaling $D(\mbf{s}) \mapsto \alpha D(\mbf{s})$ 
and $\delta x \mapsto \delta x / \alpha$, 
we can make ${\mathsf A}_{\tbox{eff}} = {\mathsf A}_{\tbox{dfr}}$
by convention.  
For special deformations $\delta x_{\tbox{NU}}$
coincides with $\delta x_c^{\tbox{qm}}$ implying 
that we get into the non-universal regime as soon 
as we have a breakdown of standard perturbation theory.

Our theoretical consideration so far do not imply 
a total collapse of perturbation theory. 
We may have in principle a co-existence of 
non-universal core component and perturbative tails. 
Actually we see such co-existence in the lower 
subfigures of Fig.7, mainly for $\delta x = 0.2512$.  
The right peak around $\kappa=0$ is clearly non-perturbative, 
while the rest of the profile is in reasonable 
(though not very good) agreement with the perturbative 
calculation. We are going to explain in the next section 
that the total collapse of perturbation theory 
for $\delta x > 0.3$  is actually not related 
at all to the appearance of non-universal features  
in the core structure. It is only circumstantial 
that in the hard wall limit this collapse happens 
as soon as we enter the non-universal regime.

\section{The collapse of perturbation theory}

A good starting for the following discussion 
is to consider the classical approximation for $P(r)$. 
Namely, in Eq.(\ref{e1}) one approximates  
$\rho_n(Q,P)$ and $\rho_m(Q,P)$ by microcanonical 
distributions. A phase-space illustration of the 
energy surfaces which support $\rho_n(Q,P)$ and $\rho_m(Q,P)$  
can be found in Fig.1 of Ref.\cite{wls}. 
The classical $P(n|m)$ equals to the overlap of these surfaces.
 
If we were dealing with a generic system 
we could introduce a linearized version of the Hamiltonian 
${\cal H}(Q,P;x) = {\cal H}_0(Q,P) + \delta x {\cal F}(Q,P;\delta x{=}0)$, 
where ${\cal F}(Q,P;x) = {\partial {\cal H}}/ {\partial x}$.
By definition this linearization is a good approximation 
provided $\delta x \ll \delta x_c^{\tbox{cl}}$, where 
$\delta x_c^{\tbox{cl}}$ is the classical correlation 
scale of ${\cal F}(Q,P;x)$ with respect to $x$. 
In the classical linear regime the classical $P(r)$ has 
the scaling property $P(r) = 1/\delta x \hat{P}(r /\delta x)$.

An equivalent definition of $\delta x_c^{\tbox{cl}}$
in the quantum-mechanical case is obtained by looking 
on the $x$ dependence of the matrix elements 
of  ${\cal F}(Q,P;x)$ in some {\em fixed~basis}. 
Again we define $\delta x_c^{\tbox{cl}}$ as the 
respective correlation scale. It is quite clear 
that for cavity with soft walls we have 
\begin{eqnarray}
\delta x_c^{\tbox{cl}} \ \ = \ \ \ell
\ \ \ \ \mbox{[soft walls]} 
\end{eqnarray}
where  $\ell=E/f$ has been defined as the penetration 
distance upon collision. 
From purely classical point of view the hard wall limit 
$\ell\rightarrow0$ is a {\em non-linear} limit. 
But this is not true quantum-mechanically. 
Here we have 
\begin{eqnarray}
\delta x_c^{\tbox{cl}} \ \ \sim \ \ \lambda_{\tbox{B}}
\ \ \ \ \mbox{[hard walls]} 
\end{eqnarray}
for $\ell < \lambda_{\tbox{B}}$. The terminology `classical 
correlation scale' while referring to $\delta x_c^{\tbox{cl}}$ 
becomes misleading here, but we shall keep using it anyway.

The theory of the core-tail structure \cite{frc} 
is valid only in the linear regime 
$\delta x \ll \delta x_c^{\tbox{cl}}$.  
Let us assume for a moment soft walls. We can ask: 
Is $\delta x \ll \delta x_c^{\tbox{cl}}$
a sufficient condition for having a core-tail structure?  
The answer is definitely not. 
Perturbation theory has a final collapse once 
the core width $b_0(\delta x)$ becomes of 
the order of the band width $b$. This defines 
a parametric scale $\delta x_{\tbox{prt}}$.  
For $\delta x > \delta x_{\tbox{prt}}$  the LDOS 
becomes purely non-perturbative. In Wigner's 
theory of random banded matrices this corresponds  
to the crossover from Lorentzian to semicircle 
line shape.

However, in the limit of hard walls the above 
mechanism of collapse becomes irrelevant  
because the band width is infinite ($b=\infty$). 
On the other hand we still have to satisfy 
the inequality $\delta x \ll \delta x_c^{\tbox{cl}}$. 
Thus, for $\delta x > \lambda_{\tbox{B}}$ we expect 
a total collapse of perturbation theory, 
as indeed observed in the numerical study.

\appendix

\section{Overlap of uncorrelated random waves}    

It is possible to estimate the overlap 
$|\langle n|m \rangle|^2$ if we assume that 
$|n\rangle$ and $|m\rangle$ are {\em uncorrelated}  
random-superpositions of plane-waves: 
A random-superpositions of plane-waves is 
characterized by the correlation function 
\begin{eqnarray} \nonumber  
\langle \psi_{\tbox{R}}(\mbf{x}_1)\psi_{\tbox{R}}(\mbf{x}_2)
\rangle \ = \ \frac{1}{{\mathsf V}} 
\ \mbox{Cos}(k|\mbf{x}_2-\mbf{x}_1|)
\end{eqnarray}
where $\mbox{Cos}(kr) \equiv 
\langle \exp(i\mbf{k}\cdot\mbf{r})\rangle_{\Omega}$ 
is a generalized Bessel function  
(for further details see Appendix D of \cite{frc}). 
Assuming that the wavefunction of $|n\rangle$ 
is uncorrelated with the wavefunction of $|m\rangle$ 
one obtains
\begin{eqnarray} \nonumber
\langle|\langle n|m \rangle|^2 \rangle \ = \ 
\left(\frac{1}{{\mathsf V}}\right)^2
\times \hspace*{4cm} \\ \nonumber
\int\!\int
\mbox{Cos}(k_n|\mbf{x}_2{-}\mbf{x}_1|) \
\mbox{Cos}(k_m|\mbf{x}_2{-}\mbf{x}_1|) \
\ d\mbf{x}_1 d\mbf{x}_2 
\end{eqnarray}
The integration is over the whole volume
of the cavity. Using the definition of the 
$\mbox{Cos}$ function we can cast this 
expression into the form 
$\langle|\langle n|m \rangle|^2 \rangle
=\left\langle f(\mbf{q}) \right\rangle_{q}$
where the average is over 
the difference $\mbf{q}=(\mbf{k}_2-\mbf{k}_1)$, 
with all possible orientations for  
$|\mbf{k}_1|=k_m$ and for $|\mbf{k}_2|=k_n$. 
The function $f(\mbf{q})$ is defined as follows:
\begin{eqnarray} \nonumber
f(\mbf{q}) \ = \ \left| \int 
\mbox{e}^{i\mbf{q}{\cdot}\mbf{x}} 
d\mbf{x} \right|^2 
\end{eqnarray}
The function $f(\mbf{q})$ depends mainly 
on $q=|\mbf{q}|$.  We can obtain an estimate 
for $f(q)$ by considering a spherical 
cavity $|\mbf{x}|<R$ with the same volume. 
Using spherical coordinates it is straightforward 
to obtain the following: 
\begin{eqnarray} \nonumber
f(q) \ \approx \ 
\left( {\cal FT} \left[ 
\frac{1}{d{-}1}(R^2-x^2)^{(d{-}1)/2}
\right]\right)^2 
\end{eqnarray}
above ${\cal FT}$ is a Fourier transform 
from $x$ to $q$. For $q \ll 1/R$ we 
have simply $f(q)=({\mathsf V})^2$. For $q \gg 1/R$
we have $f(q)=({\mathsf V})^2/(Rq)^{d{+}1}$. 
This is because of the singularity at $x=\pm R$. 
The average over $q$ can be done using again 
spherical coordinate. We have 
$q \approx ((\Delta k)^2 + 2k^2(1-\cos\theta) )^{\tbox{1/2}}$, 
where $\Delta k = |k_n-k_m|$, and we can transform 
the $d\theta$ integration into $dq$ integration:
\begin{eqnarray} \nonumber
\left\langle f(\mbf{q}) \right\rangle_{q} \ \ = \ \
\frac{\Omega_{d{-}2}}{\Omega_{d{-}1}} 
\times \hspace*{4cm} \\ \nonumber
\int \ \frac{qdq}{k^2}\left(
\frac{\sqrt{(q^2-(\Delta k)^2)((2k)^2-q^2)}}{2k^2}
\right)^{d{-}3} f(q)
\end{eqnarray}
where $\Omega_d$ is the solid angle in $d$ dimensions. 
There is no point in trying to carry an exact 
integration. Rather, it is important 
to observe that a practical approximation for 
the overlap is:
\begin{eqnarray} \nonumber 
\langle|\langle n|m \rangle|^2 \rangle \ \approx \ 
\left( \frac{1}{kR} \right)^{d{-}1} 
\frac{1}{1+(R{\cdot}(k_n{-}k_m))^2}
\end{eqnarray}
In the latter expression we have neglected 
the $d$-dependent normalization prefactor.
It is a `practical' approximation since it 
gives the correct behavior for both small 
and large values of $\Delta k$.  
The interpolation around $\Delta k \sim 1/R$  
cannot be trusted.


\ \\ \ \\ \ \\

\noindent
{\bf Acknowledgments:}

\noindent
This work was funded by ITAMP and the National Science Foundation.

\ \\ \ \\ \ \\
     



\begin{thebibliography}{99}
\bibitem{rmrk1} 
The classical approximation for $P(n|m)$ works well 
for ergodic eigenstates. As explained later in the 
text we actually consider the averaged profile. 
Namely, $P(n|m)$ is regarded as a function of $r=n-m$, 
and it is averaged over the reference state $m$.  
With this procedure the effect of the minority of non-ergodic 
eigenstates can be neglected in the classical limit.
\bibitem{rmrk2}
The two parametric scales of Wigner's RMT model 
correspond to $\delta x_c^{\tbox{qm}}$ and 
$\delta x_{\tbox{prt}}$ of Table 1. Consequently 
there are three parametric regimes in Wigner's theory. 
These are the standard perturbative regime 
($\delta x \ll \delta x_c^{\tbox{qm}}$), 
the Lorentzian regime 
($\delta x_c^{\tbox{qm}} \ll \delta x \ll \delta x_{\tbox{prt}}$), 
and the semi-circle regime ($\delta x \gg \delta x_{\tbox{prt}}$). 
As an artifact of this RMT model, there actually exists  
a forth regime (Anderson strong localization regime) 
that in the present context does not have a semiclassical analog.
\bibitem{rmrk3} 
As explained in \cite{wls} the generic hierarchy is 
$\delta x_{\tbox{NU}} \ll \delta x_{\tbox{prt}} \ll 
\delta x_{\tbox{SC}} \ll \delta x_c^{\tbox{cl}}$. 
This hierarchy is realized in the classical limit 
(small $\hbar$) where we have soft walls (see section II).
The hard wall limit is non-generic, and these four 
parametric scales coincide. In the present paper 
we assume hard walls unless stated otherwise.
\bibitem{note} 
In order to find the average profile 
we divide the $\kappa$ axis into small bins. 
We find the average $P(n|m)$ value 
in each bin. In order to get $P(r)$ 
we re-sample the average profile. The 
distance between the sampling points  
is equal to the mean level spacing. 
We are not interested in features on 
scale of the mean level spacing. Therefore 
the transformation from the continuous 
variable $\kappa$ to the integer variable 
$r$ should not be considered as problematic.
\end{thebibliography}
\end{document}